\documentclass[preprint,aps,floats,showpacs,nofootinbib]{revtex4-1}
\usepackage{amsmath,amssymb,amsfonts, epsfig}
\usepackage{graphicx}
\usepackage{epsfig}
\usepackage{amsmath}
\def\be{\begin{equation}}
\def\ee{\end{equation}}
\def\bea{\begin{eqnarray}}
\def\eea{\end{eqnarray}}
\begin{document}
\title{\bf Resurrecting Sneutrino ($\tilde{\nu}_L$) Dark Matter in light of Neutrino Mass and LUX}
\author{Arindam Chatterjee}
\email{arindam@hri.res.in}
\affiliation{Harish-Chandra Research Institute, Chhatnag Road, Jhusi, Allahabad, 211 019, India}
\author{Narendra Sahu}
\email{nsahu@iith.ac.in}
\affiliation{Department of Physics, IIT Hyderabad, ODF campus, Yeddumailaram, 502205, AP, India}
\begin{abstract}
In the minimal supersymmetric standard model (MSSM) the lightest superpartner of the 
left-handed neutrinos is ruled out of being a candidate of dark matter because of 
its large elastic cross-section with the nucleus mediated via Z-boson. We resurrect 
it by extending the MSSM with two triplets with opposite hypercharge. The addition of 
the triplets not only play a role in generating small Majorana masses for the left-handed 
active neutrinos but also make the lightest sneutrino a viable candidate for dark matter. 
We then discuss the relevant parameter space in details which can give rise to the right 
amount of (thermal)  relic abundance as well as satisfy the current direct detection 
constraints from Xenon100 and LUX. We find that sneutrino dark matter with mass 370-550 GeV 
can give rise to right thermal relic abundance while co-annihilating with the bino-like 
neutralino.   
  
\end{abstract}
\maketitle
\bigskip

\section{Introduction}\label {intr}
With the discovery of Higgs at LHC~\cite{Aad:2012tfa,Chatrchyan:2012ufa}, standard model (SM) of particle 
physics seems to be complete. However, the latter does not explain the non-zero neutrino mass, required to 
explain solar and atmospheric neutrino oscillation hypothesis, and the existence of non-baryonic dark matter (DM) 
required to explain the galaxy rotation curve, gravitational lensing and large scale structure of the 
Universe~\cite{Bertone:2004pz}. In fact, the relic abundance of DM: $\Omega_{\rm DM} h^2 \sim 0.12$, is well measured 
by WMAP-9~\cite{Hinshaw:2012aka} and Planck~\cite{Ade:2013zuv} satellites. 

The above mentioned inadequacies of SM indicate that the present form of SM is not sufficient to 
explain the current energy budget of the Universe. It needs to be extended to include sub-eV masses of 
left-handed neutrinos and the observed DM abundance. If we assume that the neutrinos are of Majorana type, 
then their sub-eV masses can be accounted through seesaw mechanisms~\cite{Minkowski:1977sc,Yanagida:1979as,
Mohapatra:1979ia,Magg:1980ut,Lazarides:1980nt,Mohapatra:1980yp,Ma:1998dx}. On the other hand, the relic abundance 
of DM can be accounted by adding an extra stable particle which is massive and electrically neutral.  

A well motivated theory beyond the SM is the minimal supersymmetric standard model (MSSM) which may explain 
DM relic abundance and sub-eV masses of left-handed neutrinos. Within MSSM, if R-parity ($R_p=(-1)^{(3B+L+2S)}$) 
is conserved, then it can easily accommodate a candidate for DM (see e.g. \cite{Drees2}). Because of conserved R-parity, 
the viable dark matter candidates are either the lightest neutralino ($\tilde{\chi}^0_1$) or the lightest left-handed sneutrino 
($\tilde{\nu}_L$). It has been known since long that $\tilde{\nu}_L$, as an elastic DM candidate, is ruled out by 
direct search limits up to a very heavy mass, beyond which it can not produce the right (thermal) relic 
abundance \cite{Falk:1994es}.  
This leaves $\tilde{\chi^0}_1$ as the only viable candidate for DM within MSSM. On the other hand, if R-parity is broken 
in MSSM then the latter does not accommodate any candidate for DM, but can explain sub-eV Majorana masses 
of light neutrinos~\cite{Borzumati:1996hd,Diaz:1997xc,Barbier:2004ez,Bhattacharyya:1996nj}. Thus a 
simultaneous explanation for sub-eV neutrino mass and DM does not exist within the framework of MSSM unless 
one adds new particles to the MSSM spectrum. 

In this article we extend the MSSM with two $SU(2)_L$ triplets~\cite{Ma:2011zm} of opposite hypercharge, such as 
$\hat{\Delta}_1 (1,3,2)$ and $\hat{\Delta}_2 (1,3,-2)$, where the numbers in the parentheses are quantum 
numbers under the gauge group $SU(3)_C \times SU(2)_L \times U(1)_Y$. We also impose a global $U(1)_{\rm B-L}$ symmetry, 
where $B$ and $L$ are baryon and lepton number respectively. Consequently all the $R$-parity violating terms in the 
MSSM superpotential are forbidden. Note that in absence of $U(1)_{\rm B-L}$ or $R$-parity, the gauge symmetry 
of MSSM superpotential allows certain terms which violate $B$ and $L$ numbers although they are strictly 
conserved within the SM. The $U(1)_{\rm B-L}$ global symmetry is allowed to be broken explicitly 
by the soft term $\Delta_1 \tilde{L} \tilde{L}$ which also breaks the supersymmetry. However, the soft term has a residual symmetry, 
$(-1)^L$ which is equivalent to a $Z_2$ symmetry. As a result the neutral candidate of $\tilde{L}$, the sneutrino, 
as the lightest supersymmetric particle (LSP), is stable. After electroweak (EW) symmetry breaking the induced vacuum 
expectation value (vev) of $\Delta_1$ generates a mass splitting between the real and imaginary parts of sneutrino. 
Assuming a mass splitting of few hundred KeV, the inelastic sneutrino (DM)-nucleon interaction mediated via $Z$ can be avoided 
\cite{Hall:1997ah,TuckerSmith:2001hy,TuckerSmith:2004jv}. A proposed explanation of DAMA~\cite{Bernabei:2013xsa} 
requires a mass splitting of $\mathcal{O} (100)$ KeV between the real and imaginary parts of $\tilde{\nu}_L$. Although 
such an explanation is disfavored by \texttt{XENON 100}~\cite{Aprile:2011ts}, a small window remains viable \cite{Arina:2012aj}. 
Moreover, small Majorana mass of neutrinos can be generated through one loop radiative process mediated by gaugino and 
sneutrino~\cite{Grossman:1997is}. 
 
Since the triplets are heavy, their CP-violating out-of-equilibrium decay can generate an asymmetry between sneutrino and 
anti-sneutrino~\cite{Ma:2011zm,Arina:2011cu} in the early Universe. However, this asymmetry can be washed out~\cite{Arina:2011cu,
Buckley:2011ye,Cirelli:2011ac} after the EW-phase transition because of sneutrino-antisneutrino mixing. Therefore, we 
focus on the parameter space where sneutrino can have the right relic abundance through thermal freeze-out 
mechanism, and at the same time sub-eV neutrino masses can be generated. Co-annihilation of sneutrino plays an important 
role in the estimation of its thermal relic abundance. In particular, co-annihilation with the bino-like neutralino and 
with the lightest sbottom (or any other strongly interacting particle) can be important in obtaining the 
right theraml relic in case of relatively light and heavy sneutrinos respectively. Typically, co-annihilation 
with the bino-like neutralino allows sneutrino masses in the range 370-550 GeV to achieve the right thermal 
relic abundance. 

The paper is arranged as follows. In section-II, we discuss the triplet extension of MSSM by focusing sneutrino as a viable 
candidate for DM and then express the relevant constraints from neutrino mass. Section-III is devoted to explain asymmetric 
sneutrino DM and its depletion through sneutrino anti-sneutrino oscillation. In section-IV, we discuss parameter space in which 
the sneutrino relic abundance can be generated through freeze-out mechanism. Section-V is devoted to discuss the constraints from 
direct detection of sneutrino DM. We conclude in section VI.

\section{Sneutrino ($\tilde{\nu}_L$) Dark Matter in Triplet Extension of MSSM}
\label{model}
We extend the MSSM superpotential by including two triplet super fields $\hat{\Delta}_1 (1,3,2)$ and $\hat{\Delta}_2 (1,3,-2)$, 
where the numbers in the parentheses are quantum numbers under the gauge group $SU(3)_C \times SU(2)_L \times U(1)_Y$. We then 
impose a global $U(1)_{\rm B-L}$ symmetry, which forbids all the $R$-parity violating terms in the MSSM superpotential. The 
relevant superpotential in presence of $U(1)_{\rm B-L}$ symmetry is given by:
\begin{equation}
\mathcal{W} \supset \mu \hat{H}_u.\hat{H}_d  + Y \hat{L}.\hat{H}_d \hat{E}^c + M \hat{\Delta}_1.\hat{\Delta}_2 
 + f_1 \hat{\Delta}_1 \hat{H}_d \hat{H}_d + f_2 \hat{\Delta}_2 \hat{H}_u \hat{H}_u \,, 
\end{equation} 
where we have suppressed the flavour indices. The corresponding Lagrangian then becomes: 
\begin{eqnarray}\label{lagrangian}
-\mathcal{L} \supset &&  |\mu|^2( |H_u|^2 + |H_d|^2 ) + M ( |\Delta_1|^2 +
|\Delta_2|^2 ) + |f_1|^2|H_d|^4 + |f_2|^2|H_u|^4 +4 |f_1|^2 |\Delta_1|^2 
|H_d|^2  \nonumber \\
&+ &  4 |f_2|^2 |\Delta_2|^2 |H_u|^2 + \left[ 2f_1^* \mu\Delta_1^\dagger H_u
H_d^\dagger + 2 f_2^* 
\mu \Delta_2^\dagger H_d H_u^\dagger + f_1^* M \Delta_2 H_d^\dagger H_d^\dagger  + f_2^* M \Delta_1 H_u^\dagger H_u^\dagger + {\rm h.c.}\right]\nonumber\\
&+& |Y|^2 \left( \tilde{L}^\dagger \tilde{L} \right) \left(H_d^\dagger H_d +  {\tilde{E^c}}^\dagger \tilde{E^c} \right) 
+ |Y|^2 \left( H_d^\dagger H_d \right) \left( {\tilde{E^c}}^\dagger \tilde{E^c} \right ) \nonumber\\
&+& \left(Y^* \mu H_u \tilde{L}^\dagger {\tilde{E^c}}^\dagger + 2 f_1 Y^* \Delta_1 H_d \tilde{L}^\dagger {\tilde{E^c}}^\dagger + {\rm h.c.} \right)\nonumber\\
&+& \mu (\tilde{H}_u. \tilde{H}_d ) + M (\tilde{\Delta}_1.\tilde{\Delta}_2) + f_1 \Delta_1 (\tilde{H}_d.\tilde{H}_d) + f_2 \Delta_2 (\tilde{H}_u.\tilde{H}_u)
+ 2 f_1 \tilde{\Delta_1} ( H_d.\tilde{H}_d) + 2 f_2 \tilde{\Delta_2} ( H_u.\tilde{H}_u) \nonumber\\
&+& Y (\tilde{L}.\tilde{H}_d) E^c + Y \tilde{E^c}(L.\tilde{H}_d) + Y (H_d.L)E^c \,. 
\end{eqnarray}
The $U(1)_{\rm B-L}$ global symmetry is explicitly broken by the soft term $\Delta_1 \tilde{L} \tilde{L}$ 
which also breaks the supersymmetry. However, the soft term has a residual symmetry, $(-1)^L$ which is equivalent 
to a $Z_2$ symmetry. As a result the neutral candidate of $\tilde{L}$, the sneutrino, can be a stable LSP. 
It will be shown later that it can be a good candidate for DM. 
In the effective theory, the relevant SUSY breaking terms in the Lagrangian are given by: 
\begin{equation}\label{soft-term}
\mathcal{V}_{\rm soft} \supset M_{\tilde{L}}^2 \tilde{L}^{*} \tilde{L}+ M B \Delta_1 \Delta_2 
+ A_1 \Delta_1 H_d H_d + A_2 \Delta_2 H_u H_u + \mu_L \Delta_1 \tilde{L} \tilde{L} + {\rm h.c.} 
\end{equation}
The co-efficient of $\Delta_1 \tilde{L} \tilde{L}$ term, i.e., $\mu_L$ is required to be small as it breaks 
$U(1)_{\rm B-L}$. The electroweak phase transition occurs when $H_u$ and $H_d$ acquire vacuum expectation values 
(vevs). They also induce small vevs for $\Delta_1$ and $\Delta_2$. From Eqs. (\ref{lagrangian}) and (\ref{soft-term}) 
we get the vevs of $\Delta_1$ and $\Delta_2$ to be
\begin{eqnarray}
\langle \Delta_1 \rangle  & \equiv & u_1 = -(A_1 v_d^2 + f_2^* M v_u^2)/2 M^2 \nonumber\\
\langle \Delta_2 \rangle  & \equiv & u_2 = -(A_2 v_u^2 + f_1^* M v_d^2)/2 M^2 \,.
\end{eqnarray} 
As we will discuss, smallness of the mass of $\nu$ requires $u_1$ to be very small. In the subsequent analysis we further 
assume $f_1$ and $f_2$ to be less than $\mathcal{O}(.1)$.  Thus the tree-level contribution to the MSSM Higgs potential 
from the triplets remain small.

\subsection{Inelastic Sneutrino Dark Matter and Constraints}
Because of the induced vevs of scalar triplets the sneutrino and anti-sneutrino states mix with each other. The 
relevant mass term takes the following form :  
\begin{equation}
 \mathcal{L}_M = \frac{1}{2} (\tilde{\nu}_L ~ \tilde{\nu}^*_L)^{*} ~~ \mathcal{M}~~
(\tilde{\nu}_L ~ \tilde{\nu}^*_L)^T,  
\end{equation}
where $\mathcal{M}$ is given by, 
\begin{equation}
\begin{pmatrix} 
M_{\tilde{L}}^2+\frac{1}{2}M_Z^2 \cos 2\beta  &  \delta M^2_{\tilde{\nu}}\cr\\
\delta M^2_{\tilde{\nu}} &  M_{\tilde{L}}^2+\frac{1}{2}M_Z^2 \cos 2\beta  
\end{pmatrix}
\end{equation}
and $\delta M^2_{\tilde{\nu}} = \mu_L u_1$. We have dropped the generation index in the above expressions. In terms 
of the CP-eigenstates $\tilde{\nu}_L=(\tilde{\nu}_{rL} + i ~\tilde{\nu}_{iL})/\sqrt{2}$. Consequently, the mass matrix 
in the basis: $(\tilde{\nu}_{rL}, \tilde{\nu}_{iL})$ is given by,
\begin{equation}
\begin{pmatrix} 
M_{\tilde{L}}^2+\frac{1}{2}M_Z^2 \cos 2\beta + \delta M^2_{\tilde{\nu}} & 0\cr\\
0 &  M_{\tilde{L}}^2+\frac{1}{2}M_Z^2 \cos 2\beta -\delta M^2_{\tilde{\nu}} 
\end{pmatrix}
\end{equation}
The eigenvalues are given by the diagonal entries and the mass eigenstates are 
given by 
\begin{equation}
\tilde{\nu}_k \in \{ \tilde{\nu}_{rL},\tilde{\nu}_{iL} \} ~\forall k \in \{1,2\}  \nonumber,
\end{equation}
where the index $k=1$ denotes the lightest state. The mass splitting between 
the two eigenvalues $\Delta M_{\tilde{\nu}} \equiv \sqrt{M_{\tilde{\nu_2}}^2-M_{\tilde{\nu_1}}^2} 
= 2~\sqrt{ |\delta M^2_{\tilde{\nu}} |} = 2 \sqrt{ |\mu_L u_1 |}$. Evading present direct detection 
bounds require $\Delta M_{\tilde{\nu}} > {\cal O} (100) {\rm KeV}$. We will come back to this issue 
in details while discussing the direct detection constraints. 

\subsection{Radiative Neutrino Mass and Constraints}
\begin{figure}
\label{figure-1}
\begin{center}
\epsfig{file=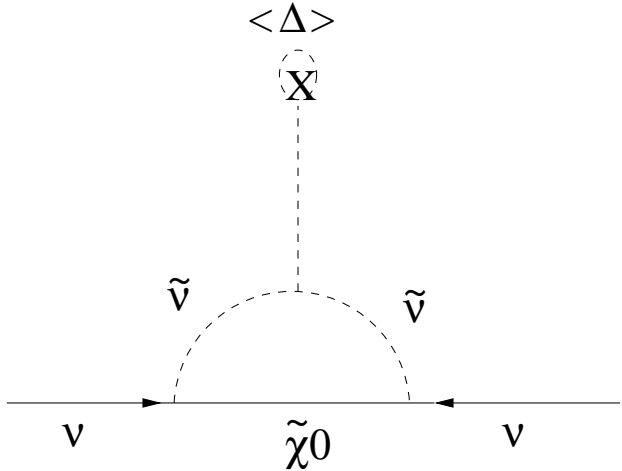, width=0.5\textwidth}
\caption{Majorana mass of neutrinos generated through one loop radiative correction.}
\end{center}
\end{figure}

\begin{figure}\label{figure-2}
\begin{center}
\epsfig{file=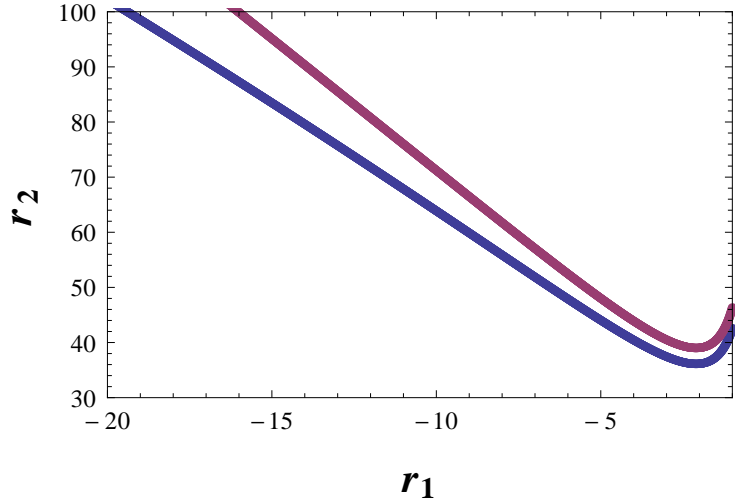,width=0.6\textwidth}
\caption{Allowed values of $r_1$ and $r_2$ are shown for all value of $ 0 < R < 10^{-5}$. 
The blue and the pink line corresponds to $R=10^{-5}$ and $R=0$ respectively. We have 
assumed a pure bino-like and a pure wino-like state with masses $|M_1|$ and $M_2$ respectively.}
\end{center}
\end{figure}

At the tree level the Majorana masses of the active neutrinos are exactly zero as we 
have imposed  an $U(1)_{\rm B-L}$ symmetry on the MSSM, which forbids not only the $R$-parity  
violating terms allowed by the MSSM superpotential, but also the $\Delta_1 L L$ term. But the $U(1)_{\rm B-L}$ 
global symmetry is softly broken to a residual symmetry $(-1)^L$ by the  $\mu_L \Delta_1 \tilde{L} \tilde{L}$. As a 
result the neutrinos acquire masses through one loop radiative correction as shown in Fig. (\ref{figure-1}). 
The neutrino mass can be calculated from Fig. (\ref{figure-1}) as \cite{Grossman:1997is,Ma:2011zm}:
\begin{equation}\label{neutrino-mass}
M_\nu= \frac{g^2}{32\pi^2 \cos \theta_w^2} \left[ \frac{ \sin \theta_w^2 r_1} {r_1^2 -1}
\left(1-\frac{r_1^2}{r_1^2-1}{\rm ln}r_1^2 \right) + \frac{\cos \theta_w^2 r_2}{r_2^2-1} \left(1-\frac{r_2^2}{r_2^2-1}{\rm ln} r_2^2 \right) 
\right] \delta M_{\tilde{\nu}}\,,
\end{equation}
where the ratios in Eq. (\ref{figure-1}) are defined by 
\begin{equation}
r_1=\frac{M_1}{M_{\tilde{\nu}}}~~~~~~{\rm and}~~~~~~~~~
r_2=\frac{M_2}{M_{\tilde{\nu}}}\,.
\end{equation} 
In the above Eq. $M_1$ and $ M_2$ are soft-supersymmetry-breaking mass parameters for 
$U(1)_Y$ and $SU(2)_L$ gauginos, which, in the limit of no-mixing, give the masses of these states. 
The non-observation of DM at direct detection experiments require 
$\Delta M_{\tilde{\nu}} > {\cal O} (100) {\rm KeV}$. On the other hand, the oscillation experiments require 
$M_\nu < 1$ eV. Thus the ratio of neutrino mass to the mass splitting of sneutrino states can be given by: 
\begin{equation}
R \equiv \frac{M_\nu}{\Delta M_{\tilde{\nu}} } < 10^{-5}\,.
\end{equation}
We have shown the allowed values of $r_1$ and $r_2$ in Fig. (\ref{figure-2}) for all 
values of $R < 10^{-5}$. For simplicity, we have assumed a pure bino--like and a pure 
wino--like neutralino with mass $|M_1|$ and $M_2$ respectively. Note that, by defining 
mass eigenstates in the neutralino-chargino sector appropriately, it is possible to 
absorb the sign of either $M_1$ or $M_2$. Thus, without loss of generality, we have 
assumed $M_2 > 0$. In order to allow for $\Delta M_{\tilde{\nu}} > {\cal O} (100) {\rm KeV}$ 
$M_1 < 0$ is required \cite{Ma:2011zm}. 

\section{Asymmetric Sneutrino Dark Matter (DM) and DM - $\overline{\rm DM}$ oscillation} 
The scalar triplets $\Delta_1$ and $\Delta_2$ are required to be heavy ($\mathcal {O}(10^{14} {\rm GeV}$)) 
in order to keep their vevs naturally small. Otherwise they will modify the $\rho$ parameter of SM. In an 
expanding Universe $\Delta_1$ and $\Delta_2$ go out-of-equilibrium as the temperature falls below their 
mass scales. As a result the CP-violating out-of-equilibrium decay of the mass eigenstates, corresponding 
to $\{ \Delta_1, \Delta_2\}$, to MSSM Higgses and sleptons can generate a net asymmetry between sleptons and 
anti-sleptons~\cite{Hambye:2000ui,Chun:2006sp,Ma:2011zm}. The asymmetry between the number densities of 
$\tilde{\nu}_L$ and $\tilde{\nu}_L^*$ 
can also be affected via the t-channel gaugino (and higgsino) mediated annihilation processes. These processes 
can annihilate a pair of $\tilde{\nu}_L$ or  $\tilde{\nu}_L^*$ producing sleptons or anti-sleptons respectively. 
However, this interaction rate is, typically, weaker than the Z-mediated s-channel process, which annihilates 
the ``symmetric'' component, i.e. annihilates one $\tilde{\nu}_L$ and with one $\tilde{\nu}_L^*$ \footnote{ There 
are t-channel neutralino mediated processes which also annihilate the ``symmetric`` component. 
However, their contribution is only secondary to the Z-mediated s-channel process.}. This reduces the total number 
density of  $\tilde{\nu}_L$ and $\tilde{\nu}_L^*$, without affecting the relative excess of $\tilde{\nu}_L$ 
compared to that of $\tilde{\nu}_L^*$. As a result one may expects a net asymmetric sneutrino dark matter.

\subsection{DM - $\overline{\rm DM}$ Oscillation and Depletion of Sneutrino Asymmetry}
After EW-phase transition the scalar triplets acquire small induced vevs. As a result the sneutrino ($\tilde{\nu}_L$) 
and anti-sneutrino ($\tilde{\nu}^*_L$) states mix with each other, thanks to the presence of $\Delta L= 2$ terms in the 
Lagrangian. This creates a small mass splitting: $\Delta M_{\tilde{\nu}} $ between the two mass eigen states: $\tilde{\nu}_{1}$ 
and $\tilde{\nu}_{2}$. The splitting between the two mass eigenstates can drive an oscillation~\cite{Arina:2011cu} as discussed 
below.
 
Let us write the sneutrino and anti-sneutrino states in terms of the mass eigenstates $\tilde{\nu}_{1}$ and 
$\tilde{\nu}_{2}$ as:
\begin{eqnarray}
|\tilde{\nu}_L \rangle  &=& \frac{1}{\sqrt{2}} \left(\tilde{\nu}_{1} + i \tilde{\nu}_{2} \right)\nonumber\\
|\tilde{\nu}^*_L \rangle  &=& \frac{1}{\sqrt{2}} \left(\tilde{\nu}_{1} - i \tilde{\nu}_{2} \right)\label{flavor-states-1}
\end{eqnarray}
The state $| \tilde{\nu}_L \rangle $ at any space-time point $(x,t)$ is given by
\be
|\phi (x,t) \rangle  = \frac{1}{\sqrt{2}} \left[ e^{-i(E_{\tilde{\nu}_{1}} t -k_{\tilde{\nu}_{1}} x)} | \tilde{\nu}_{1} \rangle 
 + i e^{+i(E_{\tilde{\nu}_{2}} t- k_{\tilde{\nu}_{2}} x)}| \tilde{\nu}_{2} \rangle \right]\,,
\label{wavefunction}
\ee
where $E_{\tilde{\nu}_{1}}=\sqrt{k_{\tilde{\nu}_{1}}^2 + M_{\tilde{\nu}_{1}}^2 }$ and $E_{\tilde{\nu}_{2}}=\sqrt{k_{\tilde{\nu}_{2}}^2 
+ M_{\tilde{\nu}_{2}}^2 }$ are the energy of $\tilde{\nu}_{1}$ and $\tilde{\nu}_{2}$ states respectively. The probability of $|\tilde{\nu}_L 
\rangle $ oscillating into $|\tilde{\nu}^*_L\rangle$ is then given by 
\be
P_{| \tilde{\nu}_L \rangle  \to | \tilde{\nu}^*_L \rangle } = |\langle \tilde{\nu}^*_L | \phi (x,t) \rangle |^2 \,.
\ee
Using Eqs.~(\ref{flavor-states-1}) and~(\ref{wavefunction}) the probability of oscillation takes the form:
\be
P_{|\tilde{\nu}_L \rangle  \to | \tilde{\nu}^*_L \rangle } = \frac{1}{4} \left[ 2 - e^{-i\left[(E_{\tilde{\nu}_{1}}-E_{\tilde{\nu}_{2}})t
- (k_{\tilde{\nu}_{2}}-k_{\tilde{\nu}_{1}})x \right]} - e^{+i\left[(E_{\tilde{\nu}_{1}}-E_{\tilde{\nu}_{2}})t - (k_{\tilde{\nu}_{2}}-k_{\tilde{\nu}_{1}})x 
\right]} \right]\,.
\label{probability}
\ee

Above the EW phase transition there is no mass splitting between the two mass eigenstates: $\tilde{\nu}_{1}$ and $\tilde{\nu}_{2}$. 
Therefore we must have $M_{\tilde{\nu}_{1}}=M_{\tilde{\nu}_{2}}$, $E_{\tilde{\nu}_{1}}=E_{\tilde{\nu}_{2}}$ and $k_{\tilde{\nu}_{1}}=k_{\tilde{\nu}_{2}}$. 
As a result from Eq.~\ref{probability} the probability of oscillation is:
\be
P_{|\tilde{\nu}_L \rangle  \to | \tilde{\nu}^*_L \rangle } = 0\,.
\ee

Below the EW phase transition the vev of $\Delta$ generates a mass splitting between the two mass eigenstates $\tilde{\nu}_{1}$ and $\tilde{\nu}_{2}$. 
Hence from Eq.~\ref{probability}, the probability of oscillation can be given by:
\be
P_{|\tilde{\nu}_L \rangle  \to |\tilde{\nu}^*_L   \rangle } \simeq \frac{1}{2} \left[ 1- \cos
\left(\frac{\Delta M_{\tilde{\nu}}^2 (t-t_{\rm EW})}{2 E_{\tilde{\nu}}} \right) \right]\,,
\label{probability_EW}
\ee
where we have assumed $E_{\tilde{\nu}_{1}} \sim E_{\tilde{\nu}_{2}} \sim E_{\tilde{\nu}}$, which is a good approximation for a small mass
splitting. In the following we will consider a mass splitting of ${\cal O}({\rm 100 keV})$. We also count 
the time of evolution from the time of EW phase transition, so that at $t=t_{\rm EW}$, $P_{|\tilde{\nu}_L \rangle  \to | \tilde{\nu}^*_L \rangle }=0$.
Below EW phase transition the time of oscillation from $\tilde{\nu}_L$ to $\tilde{\nu}^*_L$ can be estimated to be:
\be
t-t_{\rm EW} = \frac{2 E_{\tilde{\nu}} \pi}{\Delta M_{\tilde{\nu}}^2}\,.
\ee
In the relativistic limit the energy of the DM particle $E_{\tilde{\nu}} \sim T $, where $T$ is the temperature of
the thermal bath. Hence the oscillation time can be given as:
\be
t-t_{\rm EW} \sim  4 \times 10^{-14} {\rm Sec} \left( \frac{T}{100 {\rm GeV}} \right) \left( \frac{ 10^4 {\rm keV}^2}
{\Delta M_{\tilde{\nu}}^2} \right) \,.
\ee
On the other hand, in the non-relativistic limit the energy of the DM particle $E_{\tilde{\nu}} \sim M_{\tilde{\nu}}$. Thus
for $M_{\tilde{\nu}} \sim 100\  {\rm GeV}$, the time of oscillation is again similar to relativistic case. This implies that 
if the mass eigenstates $\tilde{\nu}_1$ and $\tilde{\nu}_2$ remain in the thermal equilibrium, then oscillations between these 
two states can wash out the generated asymmetry through triplet decay~\cite{Arina:2011cu,Buckley:2011ye,Cirelli:2011ac}. 
As a result we may not get any asymmetric sneutrino relic abundance. 

In order to prevent the catastrophic washout, $\tilde{\nu}$ needs to decouple from the thermal soup before the creation of 
mass splitting at EW symmetry breaking (EWSB). Assuming EWSB occurs at around 100 GeV, and considering that the freeze-out 
temperature ($T_f$) is approximately given by $\dfrac{M_{\tilde{\nu}_1}}{20}$, this requires the mass of sneutrino DM to 
be $\mathcal{O}(2 ~\text{TeV})$. However, in a scenario where, for example, if the reheat temperature after inflation is less 
than $\mathcal{O} (100~ \text{GeV})$ then this requirement may not hold good. 

If the mass of the DM, $M_{\tilde{\nu}_1}$, is less than about $\mathcal {O} (2 ~\text{TeV})$, then the initial asymmetry 
would not affect the relic density significantly. Therefore, we do not take into account the effect of any initial asymmetry 
into the present discussion. Thus, the relic density calculation resembles the case of a $\tilde{\nu}$ 
Dark Matter \cite{Hagelin:1984wv,Falk:1994es}. \footnote{The tiny mass splitting of $\cal{O}$(100) KeV between the 
states $\tilde{\nu}_1$ and $\tilde{\nu}_2$ can be ignored when these are in the thermal soup, since the freeze-out 
temperature $\cal{O}$(10) GeV is much higher compared to the splitting. The life-time of $\tilde{\nu}_2$, decaying 
to $\tilde{\nu}_1 \bar{\nu}\nu$, has been estimated to be $10^4-10^9$ seconds for a mass splitting of 100 KeV-1 MeV 
\cite{Ma:2011zm}. After freezing-out $\tilde{\nu}_2$ eventually decays to $\tilde{\nu}_1$. Also, due to very small 
decay width of $\tilde{\nu}_2$, we ignore the effect of its width in estimating the oscillation probability.}

\section{Sneutrino Dark Matter and Thermal Relic Abundance}
Assuming sufficiently high reheat temperature, and that all SUSY particles 
thermalized in the early universe, we will focus on the thermal production of 
$\tilde{\nu}_1$ Dark Matter in this section.

However, a few alterations/variations have been incorporated in the present 
discussion. Instead of expanding $\langle \sigma v_{\rm rel} \rangle$ into the leading 
$s$ and $p$ wave contributions (ignoring the higher partial waves, and 
assuming no threshold or pole in the vicinity), we have used \texttt{micrOMEGAs} 
\cite{Belanger:2004yn, Belanger:2013oya} for an accurate estimate of 
$\langle \sigma v_{rel} \rangle$, and therefore, of the relic density. 
\texttt{SuSpect} \cite{Djouadi:2002ze} has been used as the spectrum generator. 
Assuming standard cosmology, we have used the recent estimates for the 
right (thermal) relic density from the CMBR measurements by \texttt{PLANCK} 
\cite{Ade:2013zuv} and \texttt{WMAP} (9 year data)\cite{Hinshaw:2012aka}. In 
addition, we have taken into account the recent bounds on the sparticle spectrum, 
especially on the CP-even Higgs mass (125 GeV) from the LHC \cite{Chatrchyan:2012ufa,
Aad:2012tfa}. 

The computation of thermal relic abundance of the DM relies on various 
(co-)annihilation processes. This is discussed in some detail in Appendix B. 
In fact, in the absence of co-annihilations, $\Omega_{DM} h^{2} 
\propto \dfrac{1}{\langle \sigma_{ann}v \rangle} $\cite{kotu}. In the 
presence of co-annihilations, the DM and 
the co-annihilating sparticle remain in relative thermal equilibrium for 
a longer period of time through $DM~ SM \rightarrow DM'~ SM'$, where 
$DM'$ denotes the co-annihilating sparticle; $SM$ and $SM'$ denote 
two Standard Model particles, which are assumed to be in thermal equilibrium 
and therefore abundant. Of course, eventually $DM'$ decouples and decays 
to DM. Thus, co-annihilation affects the thermal relic abundance of DM. 
The effect can be captured by \cite{coann} substituting,  
\begin{equation}
 \sigma_{ann} \rightarrow \sigma_{eff} = \Sigma_{i,j} \frac{g_i g_j}{g^2_{eff}} 
(1+ \Delta_i)^{3/2} (1+ \Delta_j)^{3/2} e^{-x (\Delta_i+\Delta_j)} 
\sigma_{ij},
\label{eq:coann}
\end{equation}
where, $\{i,j\}$ runs over the list of co-annihilating sparticles, $g_i$ denotes 
the number of degrees of freedom of the $i$-th sparticle, 
$\Delta_i = \dfrac{m_{i}}{m_{DM}}-1$, $x = \dfrac{m_{DM}}{T} $ and $\sigma_{ij}$ denotes 
the co-annihilation cross-section of $i$ and $j$-th sparticles into SM particles. 
Also,  
\begin{equation}
g_{eff} = \Sigma_i g_i (1+ \Delta_i)^{3/2} e^{-x \Delta_i}.
\nonumber
\end{equation}
Thus, co-annihilations are only relevant for sufficiently small $\Delta_i$. 

Note that, there is always a (left) slepton of the same flavor as the 
$\tilde{\nu}_L$, with a small mass difference, thanks to the soft-breaking 
masses preserving the $SU(2)_L$ invariance \cite{Falk:1994es}. So, apart 
from ``co-annihilation'' with $\tilde{\nu}_2$, co-annihilation with the 
$SU(2)_L$ doublet partner will always be relevant. The dominant contributions 
to the relic abundance comes from the s-channel $Z$ mediated processes which annihilates a pair 
of $\tilde{\nu}_1, \tilde{\nu}_2$ to SM particles. The possible final states, 
for the mass-range of our interest,  are $\{f \bar{f}, W^\pm W^\mp, Z h \}$. 
Note that, due to our choice of a rather heavy $m_A$, the heavy neutral and 
charged Higgses can not occur in the final states. Also, the four point vertices 
contribute to $\{ZZ, W^\pm W^\mp, hh \}$ in final states. The processes 
with a pair of light fermion and anti-fermion in the final state are p-wave 
suppressed. So their contributions remain insignificant. Co-annihilation with 
the $SU(2)_L$ partner, via $W^\pm$ exchange also contributes.
As we will elaborate, we further include co-annihilation with various 
other sparticles, which can have significant impact on the relic density, 
opening up more parameter space where $\tilde{\nu}_1$ produces the right 
thermal relic. 

For the numerical analysis we have made the following assumptions. 

\begin{itemize}
\item For the first two generations, the squark mass parameters are 
assumed to be 2~TeV. For the 3rd generation, left (right) handed squarks 
are assumed to have soft masses around 3 (1.5) TeV. This choice alleviates 
LHC constraints from direct SUSY searches and helps to achieve the 
lightest Higgs boson mass of $\sim$ 125 GeV. The gluino mass parameter 
($M_G$) is fixed at 1.5 TeV. 

\item The soft-SUSY breaking slepton masses are assumed to be flavor-diagonal. 

\item  Trilinear soft susy breaking terms $A_t = -3.7$~TeV and $A_b = -3.7$~TeV; 
$A_{\tau} = 0$~TeV; $\tan \beta = 10$ and the CP-odd Higgs mass $m_A = 1$ TeV 
have been assumed.

\item We refrain from exact calculation of neutrino masses and mixing angles. 
$M_1 < 0 $ is assumed keeping $M_2 >0$, in order to cancel the large 
radiative contribution to the neutrino masses. $\mu = -1000 $ GeV is assumed, 
except in the context of co-annihilation with higgsino-like neutralinos. 


\item Finally we use 173.1~GeV for the top quark pole mass.
\end{itemize}

In the following we consider three scenarios in the framework of pMSSM: 
\begin{itemize}
 \item A) $\tilde{\nu}_1$ DM, with no other co-annihilating sparticles 
except the above mentioned ones; 
\item B) $\tilde{\nu}_1$ DM co-annihilating also with a bino-like neutralino 
($\tilde{\chi}_1^0$); 
\item C) $\tilde{\nu}_1$ DM co-annihilating also with a higgsino-like 
$\tilde{\chi}_1^0$ (and possibly $\tilde{\chi}_1^\pm$).  
\item D) $\tilde{\nu}_1$ DM co-annihilating also with the 
lightest $\tilde{b}$ ($\tilde{b}_1$) .  
\end{itemize}
As shown in Fig. \ref{figure-1}, since small neutrino masses require a 
-ve $M_1$ and rather large $M_2$; therefore, we refrain from discussing 
co-annihilation with a wino-like $\tilde{\chi}_1^0$ (and possibly 
$\tilde{\chi}_1^\pm$). The benchmark points are shown in table \ref{tab:bm}. 
The contribution of various (co-)annihilation channels, in each case, can 
be found in Appendix A.  

\begin{table}[t!]
\begin{center}
\begin{tabular}{|c|c|c|c|c|c| } \hline
parameter& A   & \multicolumn{2}{c|} B &  C & D  \\
\hline
  &     &  (1) & (2) &   &   \\
\hline
$M_1$                     &-1000 &-388.8  &-312.4  & -1200 & -1100\\     
$\mu$                     &-1000 &-1000   & -1000 & -677 &  -1500 \\
$m_{\tilde{L}_{3}}     $  & 580  & 385    & 310   & 690   &  1000\\
$m_{\tilde{R}}     $      & 1000 & 1000   & 1000  & 1000  &  2000\\
$m_{\tilde{\nu}_{\tau}}$  & 571.6& 379.9  & 303.6 & 687.1  &  998\\
$m_{\tilde{\nu}_{e}}$     & 998  & 496    & 303.6 & 998 &  2000\\
$m_{\tilde{\nu}_{\mu}}$   & 998  & 496    & 303.6 & 998 &  2000\\
$m_{\tilde{\chi}_1^0}$    & 962.6& 380.1  & 303.9 & 687.2 &  1090\\
\hline
$\Omega_h^2$              & 0.1  & 0.1    & 0.12 & 0.1 & 0.12 \\
\hline
\end{tabular}
\end{center}
\caption{Columns (A), (B1), (C) and (D) demonstrate benchmark points for scenarios (A), 
(B), (C) and (D) respectively. Column (B2) depicts a scenario where all three generations 
of $\tilde{\nu}$ are degenerate, and are co-annihilating with a bino-like 
$\tilde{\chi}_1^0$. All the masses are in GeV. }
\label{tab:bm}
\end{table}

In benchmark (A), we consider a scenario where $\tilde{\nu}_1$ belongs to the third 
generation, and has no additional co-annihilation channels except the above mentioned 
ones. The dominant contributions come from $\tilde{\nu}_1,~\tilde{\nu}_2$ (or $\tilde{\nu}, 
\tilde{\nu}^*$) annihilating to $Z~Z$ $(27\%)$ and $W^{\pm}~W^{\mp}$ $(24\%)$. 
While both receive contributions from 4-point vertices involving $\tilde{\nu} \tilde{\nu}^*, 
Z Z/ W^{\pm} W^{\mp} $, the $Z$ mediated $s$-channel process also 
contributes to the latter. There are $t$-channel processes mediated by $\tilde{\nu}$ 
and $\tilde{\tau}$, which also contribute to $Z Z$ and  $W^{\pm} W^{\mp}$ respectively. 
Among co-annihilation channels with (the dominantly left handed) $\tilde{\tau}_1$, 
$W^- \gamma$ and $~ W^- Z$ contribute about $9\%$ each. These processes originate from four-point 
vertices, as well as from $W$-boson exchange in the $s$-channel, while a $t$-channel contribution 
mediated by $\tilde{\tau}$ contributes sub-dominantly. The effective annihilation cross-section, as 
in eq. (\ref{eq:coann}) receives further contributions from $\tilde{\tau}_1 \tilde{\tau}_1^*$ annihilation 
into $W^{\pm}W^{\mp}$, again from the four-point vertices, and also via $s$-channel 
$Z$ exchange and $t$-channel $\tilde{\tau}$ exchange diagrams respectively. Note 
that all these dominant processes have $SU(2)_L$ gauge couplings appearing in 
the vertices.

\begin{figure}[ht!]
\begin{center}
\epsfig{file=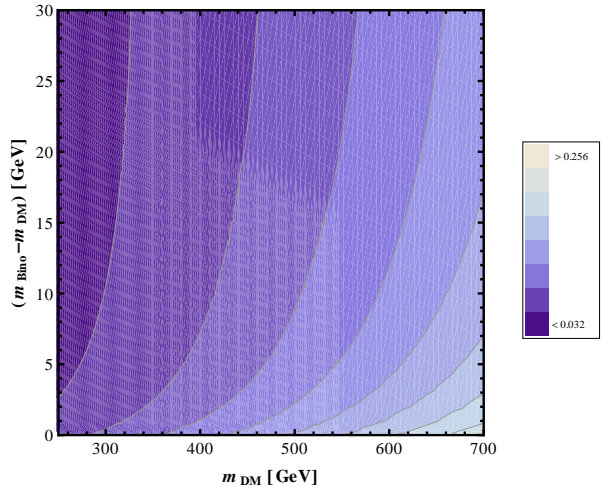, width=0.5\textwidth}
\caption{This figure shows the variation of the thermal relic density 
of $\tilde{\nu}_1$ DM, as a function of its mass and the mass difference 
with a bino-like $\chi_1^0$. }
\label{fig:coann-bino}
\end{center}
\end{figure}

In benchmark (B1), we consider further co-annihilation with a bino-like 
$\tilde{\chi}_1^0$. The mass splitting between the $\tilde{\chi}_1^0$ and 
$\tilde{\nu}_1$ (and $\tilde{\nu}_2$) is about 200 MeV. A small mass splitting 
is kept to enhance the effect of the co-annihilation. The (co-)annihilation 
processes involving a bino-like $\tilde{\chi}_1^0$ involves $U(1)_Y$ gauge 
coupling, which is less than $SU(2)_L$ gauge coupling, $\sigma_{eff}$ 
effectively becomes smaller. This contributes in a little early freeze-out 
of $\tilde{\nu}_1$ increasing the relic abundance. Thus, we get the right 
thermal relic for a lower mass of $\tilde{\nu}_1$, which is about 380 GeV. 
The dominant co-annihilation channel, in this case, is $\tilde{\chi}_1^0 \tilde{\nu}
\rightarrow W^+ \tau$ via $s$-channel $\nu$  mediation and via $t$-channel 
$\tilde{\tau}$ mediation. It contributes about 8\%. Another process 
$\tilde{\chi}_1^0 \tilde{\nu} \rightarrow Z \nu_{\tau}$, via $s$-channel $\nu$  
mediation and via 
$t$-channel $\tilde{\nu}_{\tau}$ mediation contributes about 5\%. Note that, 
since the contribution from the charge conjugate final states are also 
included, the final states for the former are twice (i.e. $ W^+ \tau$ 
and $ W^- \bar{\tau}$) that of the latter, leading to a larger contribution. 
In benchmark (B2) we consider a similar scenario, with three degenerate $\tilde{\nu}$. 
This can be achieved if the soft-mass for $SU(2)_L$ doublet sleptons are independent 
of generation. We focussed on obtaining the right thermal relic density for 
the lightest possible $\tilde{\nu}$ Dark Matter. We obtain the right thermal 
relic with sneutrino of mass $ 303.6$ GeV. We have ignored flavor mixing in the 
sneutrino (and slepton) sector. So the dominant (co-)annihilation processes 
remain the same for three generations. Figure \ref{fig:coann-bino} shows the 
variation of relic density as we vary the bino-mass parameter $M_1$. In this 
figure, we do not assume any degeneracy for all three generations of $\tilde{\nu}$. 
It demonstrates that for suitable mass difference of $\tilde{\nu}_1$ and the 
bino-like $\tilde{\chi}_1^0$, one can have the right relic density in the mass 
range of 370-550 GeV. 

In benchmark (C), we consider co-annihilation with the higgsino-like 
neutralinos and chargino. This can be achieved considering the $\mu$ parameter 
to be close to the soft-breaking mass for $\tilde{\nu}_{\tau}$. Unlike the bino, 
higgsinos come from $SU(2)_L$ doublets, and possesses relatively stronger 
interactions. Since in the limit of large $M_2$, and $|M_1| \gg \mu$ three 
states $\tilde{\chi}_1^0, ~\tilde{\chi}_2^0$ and $\tilde{\chi}_1^{\pm}$ 
are higgsino-like. Therefore, their impact on $\sigma_{eff}$ is quite 
large. The leading contribution comes from $\tilde{\chi}_1^+ ~\tilde{\chi}_1^0 
\rightarrow \{u \bar{d}, ~s \bar{c}\}$, each contributing 8\%. These occur 
dominantly via $s-$ channel $W^{\pm}$ exchange processes. Since $SU(2)_L$ 
gauge coupling appear in both vertices and because of the colour factor  
the total contribution is large. Similar $W^{\pm}$ mediated $s$-channel 
processes producing leptons contribute about 3\% each. Since $\tilde{\chi}_2^0$ 
is also higgsino-like, $\tilde{\chi}_1^+ ~\tilde{\chi}_2^0$ also annihilates to 
similar final states. However, because of the larger mass-splitting between 
$\tilde{\nu}_1$ and $\tilde{\chi}_2^0$, its contribution to $\sigma_{eff}$ is 
little less.

Finally, in benchmark (D) we consider co-annihilation with the lightest 
$\tilde{b}$, which we have assumed to be dominantly $SU(2)_L$-singlet-type. 
The mass of $\tilde{b}_1$ is assumed to be 1008.4 GeV. The dominant 
contribution to the effective thermal averaged cross-section comes from 
$ \tilde{b}_1 \tilde{b}_1^* \rightarrow g~g$; $s$-channel gluon mediation, 
as well as $t$(and $u$)-Chennai $\tilde{b}_1 $ exchange processes lead to 
this final state. This receives large enhancement due to the colour 
factor. The gluino mediated $t$-(and $u$) channel process 
$\tilde{b}_1 \tilde{b}_1 \rightarrow b b $ also contributes significantly. 
Together, these channels contribute about 80\%, as described in 
Table \ref{tab:bmD}. There are also small contributions from $\tilde{\nu} 
\tilde{\nu}^*, Z Z/ W^{\pm} W^{\mp} $.

\begin{figure}[ht!]
\label{fig:coann-bh}
\begin{center}
\epsfig{file=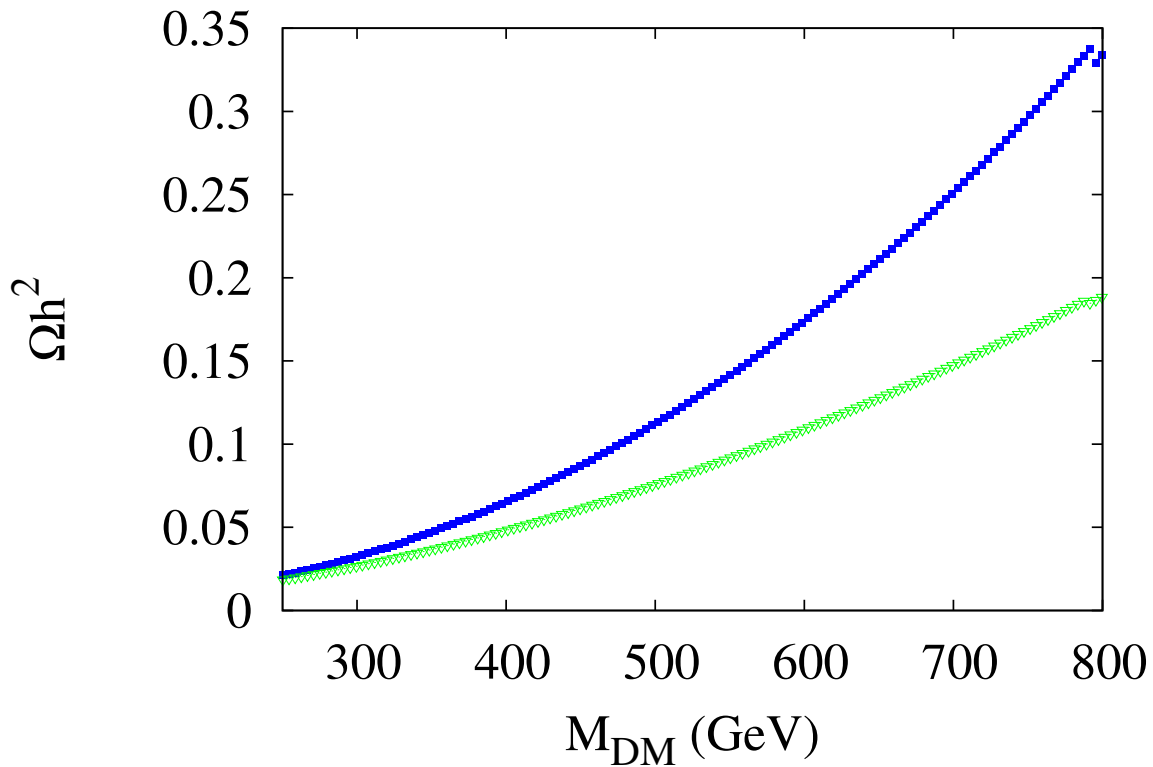, width= \textwidth}
\caption{Relic abundance for $\tilde{\nu}_1$ Dark Matter has been shown. 
The green dots represent a scenario when no additional co-annihilation 
is present; while the blue dots represent co--annihilation scenario with 
a bino--like $\tilde{\chi}^1_0$. The bino mass parameter ($|M_1|$) have been 
assumed to be 5 GeV above $\text{M}_{\text{DM}}$ in the latter scenario.} 
\end{center}
\end{figure}

In fig. \ref{fig:coann-bh}, the green line denote the thermal relic density with no 
additional co-annihilations present. Further, this figure demonstrates 
that co-annihilations with bino-like $\tilde{\chi}_1^0$ (blue line) leads 
to an increment in the relic density. We chose $|M_1| - M_{\tilde{L}} = 5 $ GeV.
 

\section{Sneutrino Dark Matter and Direct Detection Constraints}
\begin{figure}[ht!]
\begin{center}
\epsfig{file=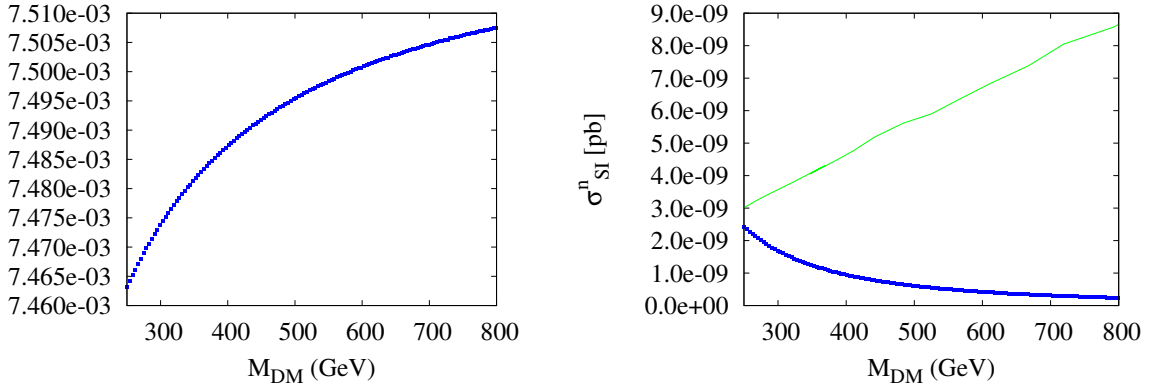, width= \textwidth}
\caption{The left panel of this figure shows the cross-section of the $\tilde{\nu}_{\tau}$ DM with neutron 
as a function of its mass. Note that this includes the $Z$ boson exchange processes. In the right panel, the 
blue line shows the Higgs exchange elastic cross-section of $\tilde{\nu}_{1}$ DM with nucleon, while the green 
line corresponds to the experimental bound from \texttt{LUX}.
}
\label{fig:dd}
\end{center}
\end{figure}

In this section we review the viability of left-handed sneutrino dark matter by taking into account 
the latest direct detection constraints from Xenon-100~\cite{Aprile:2012nq} and LUX~\cite{Akerib:2013tjd}. As 
mentioned before, the {\it dominant process} through whi ch the sneutrino DM interacts with the 
nucleon is the $t$-channel $Z$-boson mediated process, i.e. $\tilde{\nu} q \to Z \to \tilde{\nu} q$. Assuming 
sneutrino DM scattering off nucleon elastically, we have shown the DM-nucleon cross-section as a 
function of sneutrino mass in the left panel of fig. \ref{fig:dd}. From there we see that the 
cross-section is quite large and hence excludes sneutrino DM if the latter scatters off nucleon 
elastically through $t$-channel $Z$-boson mediated process. However, in the current set up, the triplet 
extension of MSSM, the sneutrino DM scatters off nucleon inelastically through the $Z$-boson 
mediated process as we have discussed in section-II. The inelastic scattering: $\tilde{\nu}_1 q \to \tilde{\nu}_2 q$ 
occurs depending on the mass splitting between the two nearly degenerate states: $\tilde{\nu}_1$ and $\tilde{\nu}_2$. 
The minimum required velocity of the sneutrino dark matter (say $\tilde{\nu}_1$) with respect the earth 
frame that will lead to a recoil inside the detector is given by: 
\begin{equation}
v_{min} = c \sqrt{\frac{1}{2 M_{\mathcal N} E_R}} \Big(\frac{M_{\mathcal N} E_R}{\mu_n}+\Delta M_{\tilde{\nu}}\Big)  \,.
\label{eq:vmin}
\end{equation}
If we assume that $\Delta M_{\tilde{\nu}}$ to be a few hundred keV, then to deposit energy inside the 
detector we need $v_{min} > v_{\rm esc}= 650 {\rm km/s}$. In other words, if the mass splitting between 
$\tilde{\nu}_1$ and $\tilde{\nu}_2$ is larger than a few hundred KeV, then sneutrino can not scatter 
off nucleon inelastically through t-channel $Z$-boson mediated process. 

The next dominant processes through which the sneutrino scatters off nucleon are the Higgs exchange 
processes occurring via the D-term. These processes receive contributions from both the CP-even Higgs 
bosons. The corresponding spin-independent cross-section, with a nucleus (N) of mass number $A$ and 
atomic number $Z$ can be expressed as, 
\begin{equation}
\sigma_0 =\frac{\mu^2}{4 \pi m_{\tilde{\nu}_1}^2 }\left(A f^p+(A-Z)f^n\right)^2\;,
\end{equation}
where $\mu = \dfrac{m_{\tilde{\nu}_1} m_{\cal{N}}}{m_{\tilde{\nu}_1}+m_{\cal{N}}}$; $m_{\cal{N}}$ 
denotes the mass of the nucleus. Further, $f_p$ and $f_n$ denotes effective 
couplings of the CP-even Higgses with proton and neutron respectively. These 
are given by, 
\begin{equation}
f^{N}= m_{N} \left( \overset{u,d,s}{\underset{q}{\Sigma}} f^{N}_{q} \frac{\lambda_q}{m_q} + 
\frac{2}{27} \overset{c,b,t}{\underset{Q}{\Sigma}} f_{G} \frac{\lambda_Q}{m_Q} \right); N \in \{p,n\}. 
\end{equation}
In the above expression $\lambda_q$ denotes the effective coupling of $\tilde{\nu}_1$ 
with the quark $q$ (i.e. ${\cal L}_{eff} \supset \lambda_q \tilde{\nu}_1^2 \bar{q} q$) in the 
limit of small momentum transfer, as is relevant for direct detection. Thus, $\lambda_q$ 
is suppressed by $m_{h/H}^2$ and is proportional to the $SU(2)_L$ gauge coupling 
$(g_2)$, the appropriate Higgs VEV and the Yukawa coupling for quark $q$ ($y_q$). $f^{N}_{q}$ denotes 
the contribution of quark $q$ to the mass $m_N$ of nucleon $N$. Note that, for large $\tan \beta$, the 
Yukawa couplings of the heavy Higgs (H) with down-type quarks can be large, and thus contributions from 
the heavy Higgs mediated channels can be significant. While the light quarks contribute to the nucleon 
masses directly, the heavy quark contributions to $f^N$ appears through the loop-induced interactions 
with gluons. These are given by, 
\begin{equation}
f^N_q = \frac{1}{m_N} \langle N |m_q \bar{q}{q}| N\rangle,
~f_G = 1 - \overset{u,d,s}{\Sigma}f^{N}_{q}.
\end{equation}
Using \texttt{micrOMEGAs-3.2}, with $\tan \beta =10$ and $m_H \simeq 500$ GeV, we estimate that the direct detection 
cross-sections fall below the present LUX bounds for the mass range of our interest. For example, a 300 GeV $\tilde{\nu}_1$ 
($\tau$-type) has a direct detection cross-section of $2\times 10^{-45} cm^2$ which is about half the limit from LUX. With 
$\tan \beta =15$ and $m_H \simeq 2000$ GeV, the interaction cross-section with neutrons have been plotted in fig. \ref{fig:dd}. 
The cross-section with protons is also similar.  
Note that, the strange quark content of the nucleon has significant uncertainties, 
leading to an uncertainty in $f^N_s$. In the Higgs mediated processes, the $s$-quark content 
plays an important role, due to its large Yukawa coupling. We have used the default values 
in \texttt{micrOMEGAs-3.2} to estimate the cross-section. Note that by varying $f^N_s$ 
it is possible to reduce the direct detection cross-section even further.

\section{Conclusion}
We discussed the viability of left-handed sneutrino ($\tilde{\nu}_L$) as a candidate for DM
in the triplet extension of the minimal supersymmetric standard model (MSSM). We extended the MSSM with two
triplets of opposite hypercharges and imposed a global $U(1)_{\rm B-L}$ symmetry. The $B-L$ symmetry is then
allowed to break explicitly by a $\Delta L=2$ term ($\Delta_1 \tilde{L} \tilde{L}$) which has a residual $Z_2$
symmetry. As a result the lightest left-handed sneutrino became stable and a viable candidate for DM. 
It is worth mentioning that within MSSM, sneutrino is ruled out as a candidate for (elastic) DM because 
of its large direct detection cross-section with the nucleus mediated via Z-boson. However, in the 
triplet extension of MSSM, this problem has been eradicated by creating a mass splitting between the 
real and imaginary parts of the sneutrino DM. By choosing the mass splitting to be a few hundred KeV, 
the $Z$-mediated process $\tilde{\nu}_1 q \to Z\to \tilde{\nu}_1 q $ is forbidden. We then 
discussed the elastic scattering of sneutrino DM with the nucleon via the Higgs exchange processes. In fact, we found 
that for a 300 GeV sneutrino DM mass, the DM-nucleon cross-section is approximately $2\times 10^{-45} {\rm cm}^2$ which is 
about half the limit from LUX.    

Assuming that sneutrino is in thermal equilibrium in the early Universe, we estimated its relic density. We showed
that, in a large part of the parameter space, co-annihilation of sneutrino plays an important role in the relic 
abundance estimation. In particular, assuming that mass splitting with the bino-like neutralino is small, we showed 
that the allowed mass of DM is in the range of 370-550 GeV. Note that for such range of sneutrino mass, the LUX bound 
is completely evaded.

Since the lepton number is broken explicitly by two units, the Majorana masses of light neutrinos could be generated 
at loop level. Further, since the additional triplets are very heavy, the model resembles MSSM in the energy 
accessible to the LHC. In summary, the salient features of this scenario include a very heavy wino and the possibility 
of having a $\tilde{\nu}_L$-type LSP which is a suitable candidate for DM. In future we will explore its collider 
phenomenology in detail.

\section*{Acknowledgement} We would like to thank the organizers of the Workshop on High Energy Physics and Phenomenology
(WHEPP13), held at Puri, Odisha, India during 12-21 December 2013 where the foundation of this work was laid. We would 
also like to thank Debottam Das for useful discussions during the initial stage. NS is partially
supported by the Department of Science and Technology Grant SR/FTP/PS-209/2011.

\section*{Appendix A}
In this appendix, we sketch the thermal relic density calculation\cite{kotu, Edsjo:1997bg}. 
We assume that, to begin with, all the sparticles and the SM particles were 
in thermal equilibrium, forming a thermal soup. However, as the expansion rate
of the universe exceeds the interaction rate (of the interactions which kept the 
species in thermal equilibrium) of a particle species, they decouple from the thermal 
soup. Due to the conserved R-parity, the total number of sparticles (in the early universe) 
is reduced only by their annihilation into the SM particles. Therefore the relevant 
number density to consider, to begin with, is the number density of all the sparticles 
(say $n$), since the remaining ones (not annihilating into the SM particles) will 
decay to the $\tilde{\nu}_1$ eventually contributing to the number density of 
$\tilde{\nu}_1$. The Boltzmann equation, governing the the evolution of the 
number density $n$, (in the FRW background, see e.g. \cite{kotu}) can be written as, 
\begin{equation}
\frac{d n }{dt} + 3 H n = -\langle \sigma v \rangle (n^{2}- n ^{2}_{eq}), 
 \end{equation}
where $n^{2}_{eq}$ denotes the equilibrium abundance. In this equation the 
second term in the left hand side arises due the expansion of the universe, 
the Hubble parameter being denoted by $H$. To scale out the effect due to the 
expansion of the universe, one often uses $\dfrac{n_{\tilde{\nu}_1}}{s}$, where $s$ 
denotes the entropy density to rewrite the above equation as, 
\begin{equation}
 \frac{dY}{dT}= \sqrt{\frac{\pi  g_*(T) }{45}} M_p \langle \sigma v \rangle 
 (Y(T)^2-Y_{eq}(T)^2),
\label{relicd}
\end{equation}
where $T$ stands for the temperature, $Y = \dfrac{n_{\tilde{\nu}_1}}{\text{s}}$, $g_{*}$ is an 
effective number relativistic degrees of freedom  and $M_p$ is the Planck mass. 
In order to express the time derivative in terms of the temperature derivative, conservation
of the comoving entropy has been used, which gives $ \frac{dT}{dt} = -H$, where 
$H$, as mentioned already, is the Hubble parameter. Note that, therefore, late \footnote{ 
    By late we mean after the freeze--out of $\tilde{\nu}_1$, i.e. after the rate of the reaction 
    pair producing $\tilde{\nu}_1$ falls behind the Hubble expansion rate (given by the Hubble 
    parameter $H$).} production of entropy (although not possible in the present scenario) will 
alter this discussion, see e.g. \cite{kotu, Drees2}. 
Further, $\langle \sigma v \rangle $ represents the relativistic thermally averaged 
(effective) annihilation cross-section of superparticles (into the SM particles) and 
is expressed as, \footnote{Since the freeze--out of the species under consideration 
  occurs at a  temperature well below its mass, typically $T_F \sim m/ 20$, 
  Maxwell--Boltzmann statistics have been used in eq.(\ref{sv}).}.  
\begin{equation}
       \langle \sigma v \rangle=  \frac{ \sum\limits_{i,j}g_i g_j  \int\limits_{(m_i+m_j)^2} ds\sqrt{s}
K_1(\sqrt{s}/T) p_{ij}^2\sigma_{ij}(s)}{2T\big(\sum\limits_i g_i m_i^2 K_2(m_i/T)\big)^2 }\;. 
\label{sv}
\end{equation}

In eq.(\ref{sv}), the sum over $i\, , j$ spans over all the sparticles, 
$m_i$ denote the mass of sparticle (labeled by) $i$, $\sigma_{ij}$ denotes the annihilation 
cross-section of the sparticles $i$ and $j$ into the SM particles, $p_{ij}$ and $\sqrt{s}$ 
   \footnote{Note that we have also used ``s'' for the entropy density.} denote the momentum 
and the total energy of the ``incoming'' sparticles in their center-of-mass frame. $K_1$ and 
$K_2$ denote modified Bessel functions of type one and two respectively. 
Eq.(\ref{relicd}) can not be solved exactly by analytical means, a discussion on 
approximate solution can be found, e.g., in \cite{kotu}. However, we have used the publicly 
available code \texttt{micrOMEGAs} for the relic density calculation which solves 
eq.(\ref{sv}) numerically without any approximation. For thermal averaging, we 
consider only sparticles ($i\, ,j$) such that the Boltzmann suppression factor 
$e^{-A}$, where $A=\left(\dfrac{m_i+m_j -2 m_{\tilde{\nu}_1}}{T}\right)$ less than $10^{-6}$, 
which is (more than) sufficient for our purpose.  

Solving eq.(\ref{sv}) by integrating over $T_F$ to $T_0$, $T_0$ being the present 
CMBR temperature, gives the present value of $Y$, which we denote by $Y_0$. The 
present relic density, then, as a fraction of the critical density (which corresponds 
to a ``flat'' universe) can be expressed as, 
\begin{equation}
\Omega_{0 \tilde{\nu}_1}h^2  =  \frac{\rho_{\tilde{\nu}_1}}{\rho_{crit}} h^2 
= \frac{m_{\tilde{\nu}_1} s_0 Y_0}{\rho_{crit}} h^2 
 = 2.742 \times m_{\tilde{\nu}_1} Y_0 /\text{GeV}.
\end{equation}

where $\rho_{crit}= \dfrac{3 H_{0}^2}{8 \pi G}$, with $H_0$ and $G$ denoting the 
(present) Hubble's parameter and the gravitational constant respectively and $s_0$  
denotes the present entropy density.  

\section*{Appendix B}
In this Appendix, we mention the (co-)annihilation channels which 
contribute more than 1\% to the relic density, in case of the benchmark 
points shown in Table \ref{tab:bm}. We obtain these estimates from 
\texttt{micrOMEGAs}. Because of the tiny mass splitting between 
$\tilde{\nu}_1$ and $\tilde{\nu}_2$, we simply use $\tilde{\nu}$ and 
$\tilde{\nu}^*$ instead in the following tables. Note that we have 
ignored any flavor oscillation in the sneutrino sector. While introducing the 
flavor oscillations will not affect the relic density in a significant manner; 
it will affect the relative contributions from flavor dependent final 
states. We mention all relevant channels contributing more that $1\%$ to the 
relic density.

\begin{table}[!ht]
\begin{center}
\begin{tabular}{|c|c|c| } \hline
Initial states & Final states &  Contribution to $\Omega_{DM}^{-1}$ (in \%)  
\\ \hline
$\tilde{\nu}_{\tau}$ $\tilde{\nu}_{\tau}^*$ & $W^+$ $W^-$ & 27 \\     
$\tilde{\nu}_{\tau}$ $\tilde{\nu}_{\tau}^*$ & $Z$ $Z$   & 24 \\
$ \tilde{\tau}_1$ $\tilde{\nu}_{\tau}^*$ & $\gamma$ $W^-$  & 10 \\
$\tilde{\tau}_1$ $\tilde{\tau}_1^*$ & $W^+$ $W^-$ & 9  \\
$\tilde{\tau}_1$ $\tilde{\nu}_{\tau}^*$ & Z $W^-$  & 9  \\
$\tilde{\tau}_1$ $\tilde{\nu}_{\tau}^*$ & $W^-$ h  & 6 \\
$\tilde{\tau}_1$ $\tilde{\tau}_1^*$ &\{$\gamma$ $\gamma$,h h\}& 3 \\
$\tilde{\tau}_1$ $\tilde{\tau}_1^*$ & $\gamma$ $Z$   & 2 \\
$\tilde{\tau}_1$ $\tilde{\nu}_{\tau}^*$ & $\bar{t}$ $b$   & 1 \\
\hline
\end{tabular}
\end{center}
\caption{Contribution from various annihilation and co-annihilation channels
to the relic density of $\tilde{\nu}_1$ Dark Matter, for benchmark (A) 
of table \ref{tab:bm}. }
\label{tab:bmA}
\end{table}
\begin{table}[!ht]
\begin{center}
\begin{tabular}{|c|c|c| } \hline
Initial states & Final states &  Contribution to $\Omega_{DM}^{-1}$ (in \%)  
\\ \hline
$\tilde{\nu}_{\tau}$ $\tilde{\nu}_{\tau}^*$ & $W^+$ $W^-$ & 28 \\     
$\tilde{\nu}_{\tau}$ $\tilde{\nu}_{\tau}^*$ & $Z$ $Z$   & 24 \\
$\tilde{\chi}_1^0$ $\tilde{\nu}_{\tau}$ & $W^+$ $\tau$  & 8  \\
$\tilde{\tau}_1$ $\tilde{\nu}_{\tau}^*$ & $\gamma$ $W^-$  & 6  \\
$\tilde{\tau}_1$ $\tilde{\nu}_{\tau}^*$ & $Z$ $W^-$  & 5  \\
$\tilde{\chi}_1^0$ $\tilde{\nu}_{\tau}$ & $Z$ $\nu_{\tau}$  & 5 \\
$\tilde{\tau}_1$ $\tilde{\tau}_1^*$ & $W^+$ $W^-$ & 4 \\
$\tilde{\nu}_{\tau}$ $\tilde{\nu}_{\tau}^*$ & $h$ $h$   & 3 \\
$\tilde{\tau}_1$ $\tilde{\nu}_{\tau}^*$ & $W^-$ $h$  & 2 \\
$\tilde{\nu}_{\tau}$ $\tilde{\nu}_{\tau}$ & $\nu_{\tau}$ $\nu_{\tau}$ & 1 \\
$\tilde{\tau}_1$ $\tilde{\tau}_1^*$ & \{$\gamma$ $\gamma$, $\gamma$ Z\} & 1\\
$\tilde{\nu}_{\tau}$ $\tilde{\nu}_{\tau}^*$ & $t$ $\bar{t}$ & 1\\
$\tilde{\tau}_1$ $\tilde{\nu}_{\tau}^*$ & $\bar{t}$ $b$ & 1\\
\hline
\end{tabular}
\end{center}
\caption{Contribution from various annihilation and co-annihilation channels
to the relic density of $\tilde{\nu}_1$ Dark Matter, for benchmark (B1) 
of table \ref{tab:bm}. }
\label{tab:bmB1}
\end{table}
\begin{table}[!ht]
\begin{center}
\begin{tabular}{|c|c|c| } \hline
Initial states & Final states &  Contribution to $\Omega_{DM}^{-1}$ (in \%)  
\\ \hline
$\tilde{\nu}_{i}$ $\tilde{\nu}_{i}^*$ & $W^+$ $W^-$ & 30 \\
$\tilde{\nu}_{i}$ $\tilde{\nu}_{i}^*$ & $Z$ $Z$   & 27 \\
$\tilde{\chi}_1^0$ $\tilde{\nu}_{i}$ & $W^+$ $l_i$ & 9 \\
$\tilde{\chi}_1^0$ $\tilde{\nu}_{i}$ & $Z$ $\nu_i$  & 6 \\
$\tilde{\nu}_{i}$ $\tilde{\nu}_{i}^*$ & $h$ $h$   & 3 \\
$\tilde{l}_i$ $\tilde{\nu}_{i}^*$ & $\gamma$ $W^-$  & 3 \\
$\tilde{l}_i$ $\tilde{\nu}_{i}^*$ & $Z$ $W^-$  & 3 \\
\hline
\end{tabular}
\end{center}
\caption{Contribution from various annihilation and co-annihilation channels
to the relic density of $\tilde{\nu}_1$ Dark Matter, for benchmark (B2) 
of table \ref{tab:bm}. The subscript $i \in \{1,2,3\}$ denotes generation. }
\label{tab:bmB2}
\end{table}

\begin{table}[!ht]
\begin{center}
\begin{tabular}{|c|c|c| } \hline
Initial states & Final states &  Contribution to $\Omega_{DM}^{-1}$ (in \%)  
\\ \hline
$\tilde{b}_{1}$ $\tilde{b}_{1}^*$ & $g$ $g$ & 58 \\
$\tilde{b}_{1}$ $\tilde{b}_{1}$ & $b$ $b$   & 19 \\
$\tilde{\nu}_{\tau}$ $\tilde{\nu}_{\tau}^*$ & $W^+$ $W^-$ & 3 \\     
$\tilde{\nu}_{\tau}$ $\tilde{\nu}_{\tau}^*$ & $Z$ $Z$   & 3 \\
$\tilde{b}_{1}$ $\tilde{b}_{1}^*$ & $\gamma$ $g$   & 2 \\
$\tilde{\tau}_1$ $\tilde{\nu}_{\tau}^*$ & $\gamma$ $W^-$ & 2 \\
$\tilde{\tau}_1$ $\tilde{\tau}_1^*$ & $W^+$ $W^-$  & 1 \\
$\tilde{\tau}_1$ $\tilde{\nu}_{\tau}^*$ & $Z$ $W^-$ & 1 \\
\hline
\end{tabular}
\end{center}
\caption{Contribution from various annihilation and co-annihilation channels
to the relic density of $\tilde{\nu}_1$ Dark Matter, for benchmark (D) 
of table \ref{tab:bm}. The subscript $i \in \{1,2,3\}$ denotes generation. }
\label{tab:bmD}
\end{table}

\begin{table}[!ht]
\begin{center}
\begin{tabular}{|c|c|c| } \hline
Initial states & Final states &  Contribution to $\Omega_{DM}^{-1}$ (in \%)  
\\ \hline
$\tilde{\chi}_1^+$ $\tilde{\chi}_1^0$ & \{$u$ $\bar{d}$, $\bar{s}$ $c$\}  & 8  \\
$\tilde{\chi}_1^+$ $\tilde{\chi}_2^0$ & \{$u$ $\bar{d}$, $\bar{s}$ $c$\}  & 5\\    
$\tilde{\nu}_{\tau}$ $\tilde{\nu}_{\tau}^*$ & \{$W^+$ $W^-$, $Z$ $Z$\}& 3\\
$\tilde{\chi}_1^0$ $\tilde{\chi}_1^0$ & $W^+$ $W^-$       & 3\\
$\tilde{\chi}_1^+$ $\tilde{\chi}_1^0$ & \{$\nu_e$ $\bar{e}$, $\nu_{\mu}$ $\bar{m}$, $\nu_{\tau}$ $\bar{l}$\}& 3\\
$\tilde{\chi}_1^0$ $\tilde{\chi}_2^0$ & \{$d$ $\bar{d}$, $s$ $\bar{s}$, $b$ $\bar{b}$, $u$ $\bar{u}$, $c$ $\bar{c}$\}& 2\\
$\tilde{\chi}_1^0$ $\tilde{\chi}_1^0$ & $Z$ $Z$         & 2\\
$\tilde{\chi}_1^+$ $\tilde{\chi}_1^-$ & \{$W^+$ $W^-$, $u$ $\bar{u}$, $c$ $\bar{c}$, $t$ $\bar{t}$\}& 2\\
$\tilde{\chi}_1^+$ $\tilde{\chi}_1^0$ & $t$ $\bar{b}$         & 2\\
$\tilde{\chi}_1^+$ $\tilde{\chi}_2^0$ & \{$\nu_e$ $\bar{e}$, $\nu_{\mu}$ $\bar{m}$, $\nu_{\tau}$ $\bar{l}$\}& 2\\
$\tilde{\chi}_1^+$ $\tilde{\chi}_2^0$ & $t$ $\bar{b}$         & 1\\
$\tilde{\chi}_1^+$ $\tilde{\chi}_1^0$ & $\gamma$ $W^+$        & 1\\
$\tilde{\chi}_1^+$ $\tilde{\chi}_1^-$ & \{$d$ $\bar{d}$, s $\bar{s}$\}  & 1\\ 
$\tilde{\tau}_1$ $\tilde{\nu}_{\tau}^*$ & $\gamma$ $W^-$        & 1\\
$\tilde{\chi}_1^0$ $\tilde{\chi}_2^0$ & \{$\nu_e$ $\bar{\nu}_e$, $\nu_{\mu}$ $\bar{\nu}_{\mu}$, $\nu_{\tau}$ $\bar{\nu}_{\tau}$\} & 1\\
$\tilde{\tau}_1$ $\tilde{\tau}_1^*$ & $W^+$ $W^-$       & 1\\
\hline
\end{tabular}
\end{center}
\caption{Contribution from various annihilation and co-annihilation channels
to the relic density of $\tilde{\nu}_1$ Dark Matter, for benchmark (C) 
of table \ref{tab:bm}. }
\label{tab:bmC}
\end{table}

\clearpage
\bibliographystyle{apsrev4-1}
\bibliography{sneutrinodm}

\begin{thebibliography}{40}%
\makeatletter
\providecommand \@ifxundefined [1]{%
 \@ifx{#1\undefined}
}%
\providecommand \@ifnum [1]{%
 \ifnum #1\expandafter \@firstoftwo
 \else \expandafter \@secondoftwo
 \fi
}%
\providecommand \@ifx [1]{%
 \ifx #1\expandafter \@firstoftwo
 \else \expandafter \@secondoftwo
 \fi
}%
\providecommand \natexlab [1]{#1}%
\providecommand \enquote  [1]{``#1''}%
\providecommand \bibnamefont  [1]{#1}%
\providecommand \bibfnamefont [1]{#1}%
\providecommand \citenamefont [1]{#1}%
\providecommand \href@noop [0]{\@secondoftwo}%
\providecommand \href [0]{\begingroup \@sanitize@url \@href}%
\providecommand \@href[1]{\@@startlink{#1}\@@href}%
\providecommand \@@href[1]{\endgroup#1\@@endlink}%
\providecommand \@sanitize@url [0]{\catcode `\\12\catcode `\$12\catcode
  `\&12\catcode `\#12\catcode `\^12\catcode `\_12\catcode `\%12\relax}%
\providecommand \@@startlink[1]{}%
\providecommand \@@endlink[0]{}%
\providecommand \url  [0]{\begingroup\@sanitize@url \@url }%
\providecommand \@url [1]{\endgroup\@href {#1}{\urlprefix }}%
\providecommand \urlprefix  [0]{URL }%
\providecommand \Eprint [0]{\href }%
\providecommand \doibase [0]{http://dx.doi.org/}%
\providecommand \selectlanguage [0]{\@gobble}%
\providecommand \bibinfo  [0]{\@secondoftwo}%
\providecommand \bibfield  [0]{\@secondoftwo}%
\providecommand \translation [1]{[#1]}%
\providecommand \BibitemOpen [0]{}%
\providecommand \bibitemStop [0]{}%
\providecommand \bibitemNoStop [0]{.\EOS\space}%
\providecommand \EOS [0]{\spacefactor3000\relax}%
\providecommand \BibitemShut  [1]{\csname bibitem#1\endcsname}%
\let\auto@bib@innerbib\@empty
\bibitem [{\citenamefont {Aad}\ \emph {et~al.}(2012)\citenamefont {Aad} \emph
  {et~al.}}]{Aad:2012tfa}%
  \BibitemOpen
  \bibfield  {author} {\bibinfo {author} {\bibfnamefont {G.}~\bibnamefont
  {Aad}} \emph {et~al.} (\bibinfo {collaboration} {ATLAS Collaboration}),\
  }\href {\doibase 10.1016/j.physletb.2012.08.020} {\bibfield  {journal}
  {\bibinfo  {journal} {Phys.Lett.}\ }\textbf {\bibinfo {volume} {B716}},\
  \bibinfo {pages} {1} (\bibinfo {year} {2012})},\ \Eprint
  {http://arxiv.org/abs/1207.7214} {arXiv:1207.7214 [hep-ex]} \BibitemShut
  {NoStop}%
\bibitem [{\citenamefont {Chatrchyan}\ \emph {et~al.}(2012)\citenamefont
  {Chatrchyan} \emph {et~al.}}]{Chatrchyan:2012ufa}%
  \BibitemOpen
  \bibfield  {author} {\bibinfo {author} {\bibfnamefont {S.}~\bibnamefont
  {Chatrchyan}} \emph {et~al.} (\bibinfo {collaboration} {CMS Collaboration}),\
  }\href {\doibase 10.1016/j.physletb.2012.08.021} {\bibfield  {journal}
  {\bibinfo  {journal} {Phys.Lett.}\ }\textbf {\bibinfo {volume} {B716}},\
  \bibinfo {pages} {30} (\bibinfo {year} {2012})},\ \Eprint
  {http://arxiv.org/abs/1207.7235} {arXiv:1207.7235 [hep-ex]} \BibitemShut
  {NoStop}%
\bibitem [{\citenamefont {Bertone}\ \emph {et~al.}(2005)\citenamefont
  {Bertone}, \citenamefont {Hooper},\ and\ \citenamefont
  {Silk}}]{Bertone:2004pz}%
  \BibitemOpen
  \bibfield  {author} {\bibinfo {author} {\bibfnamefont {G.}~\bibnamefont
  {Bertone}}, \bibinfo {author} {\bibfnamefont {D.}~\bibnamefont {Hooper}}, \
  and\ \bibinfo {author} {\bibfnamefont {J.}~\bibnamefont {Silk}},\ }\href
  {\doibase 10.1016/j.physrep.2004.08.031} {\bibfield  {journal} {\bibinfo
  {journal} {Phys.Rept.}\ }\textbf {\bibinfo {volume} {405}},\ \bibinfo {pages}
  {279} (\bibinfo {year} {2005})},\ \Eprint
  {http://arxiv.org/abs/hep-ph/0404175} {arXiv:hep-ph/0404175 [hep-ph]}
  \BibitemShut {NoStop}%
\bibitem [{\citenamefont {Hinshaw}\ \emph {et~al.}(2013)\citenamefont {Hinshaw}
  \emph {et~al.}}]{Hinshaw:2012aka}%
  \BibitemOpen
  \bibfield  {author} {\bibinfo {author} {\bibfnamefont {G.}~\bibnamefont
  {Hinshaw}} \emph {et~al.} (\bibinfo {collaboration} {WMAP}),\ }\href
  {\doibase 10.1088/0067-0049/208/2/19} {\bibfield  {journal} {\bibinfo
  {journal} {Astrophys.J.Suppl.}\ }\textbf {\bibinfo {volume} {208}},\ \bibinfo
  {pages} {19} (\bibinfo {year} {2013})},\ \Eprint
  {http://arxiv.org/abs/1212.5226} {arXiv:1212.5226 [astro-ph.CO]} \BibitemShut
  {NoStop}%
\bibitem [{\citenamefont {Ade}\ \emph {et~al.}(2013)\citenamefont {Ade} \emph
  {et~al.}}]{Ade:2013zuv}%
  \BibitemOpen
  \bibfield  {author} {\bibinfo {author} {\bibfnamefont {P.}~\bibnamefont
  {Ade}} \emph {et~al.} (\bibinfo {collaboration} {Planck Collaboration}),\
  }\href@noop {} {\  (\bibinfo {year} {2013})},\ \Eprint
  {http://arxiv.org/abs/1303.5076} {arXiv:1303.5076 [astro-ph.CO]} \BibitemShut
  {NoStop}%
\bibitem [{\citenamefont {Minkowski}(1977)}]{Minkowski:1977sc}%
  \BibitemOpen
  \bibfield  {author} {\bibinfo {author} {\bibfnamefont {P.}~\bibnamefont
  {Minkowski}},\ }\href {\doibase 10.1016/0370-2693(77)90435-X} {\bibfield
  {journal} {\bibinfo  {journal} {Phys.Lett.}\ }\textbf {\bibinfo {volume}
  {B67}},\ \bibinfo {pages} {421} (\bibinfo {year} {1977})}\BibitemShut
  {NoStop}%
\bibitem [{\citenamefont {Yanagida}(1979)}]{Yanagida:1979as}%
  \BibitemOpen
  \bibfield  {author} {\bibinfo {author} {\bibfnamefont {T.}~\bibnamefont
  {Yanagida}},\ }\href@noop {} {\bibfield  {journal} {\bibinfo  {journal}
  {Conf.Proc.}\ }\textbf {\bibinfo {volume} {C7902131}},\ \bibinfo {pages} {95}
  (\bibinfo {year} {1979})}\BibitemShut {NoStop}%
\bibitem [{\citenamefont {Mohapatra}\ and\ \citenamefont
  {Senjanovic}(1980)}]{Mohapatra:1979ia}%
  \BibitemOpen
  \bibfield  {author} {\bibinfo {author} {\bibfnamefont {R.~N.}\ \bibnamefont
  {Mohapatra}}\ and\ \bibinfo {author} {\bibfnamefont {G.}~\bibnamefont
  {Senjanovic}},\ }\href {\doibase 10.1103/PhysRevLett.44.912} {\bibfield
  {journal} {\bibinfo  {journal} {Phys.Rev.Lett.}\ }\textbf {\bibinfo {volume}
  {44}},\ \bibinfo {pages} {912} (\bibinfo {year} {1980})}\BibitemShut
  {NoStop}%
\bibitem [{\citenamefont {Magg}\ and\ \citenamefont
  {Wetterich}(1980)}]{Magg:1980ut}%
  \BibitemOpen
  \bibfield  {author} {\bibinfo {author} {\bibfnamefont {M.}~\bibnamefont
  {Magg}}\ and\ \bibinfo {author} {\bibfnamefont {C.}~\bibnamefont
  {Wetterich}},\ }\href {\doibase 10.1016/0370-2693(80)90825-4} {\bibfield
  {journal} {\bibinfo  {journal} {Phys.Lett.}\ }\textbf {\bibinfo {volume}
  {B94}},\ \bibinfo {pages} {61} (\bibinfo {year} {1980})}\BibitemShut
  {NoStop}%
\bibitem [{\citenamefont {Lazarides}\ \emph {et~al.}(1981)\citenamefont
  {Lazarides}, \citenamefont {Shafi},\ and\ \citenamefont
  {Wetterich}}]{Lazarides:1980nt}%
  \BibitemOpen
  \bibfield  {author} {\bibinfo {author} {\bibfnamefont {G.}~\bibnamefont
  {Lazarides}}, \bibinfo {author} {\bibfnamefont {Q.}~\bibnamefont {Shafi}}, \
  and\ \bibinfo {author} {\bibfnamefont {C.}~\bibnamefont {Wetterich}},\ }\href
  {\doibase 10.1016/0550-3213(81)90354-0} {\bibfield  {journal} {\bibinfo
  {journal} {Nucl.Phys.}\ }\textbf {\bibinfo {volume} {B181}},\ \bibinfo
  {pages} {287} (\bibinfo {year} {1981})}\BibitemShut {NoStop}%
\bibitem [{\citenamefont {Mohapatra}\ and\ \citenamefont
  {Senjanovic}(1981)}]{Mohapatra:1980yp}%
  \BibitemOpen
  \bibfield  {author} {\bibinfo {author} {\bibfnamefont {R.~N.}\ \bibnamefont
  {Mohapatra}}\ and\ \bibinfo {author} {\bibfnamefont {G.}~\bibnamefont
  {Senjanovic}},\ }\href {\doibase 10.1103/PhysRevD.23.165} {\bibfield
  {journal} {\bibinfo  {journal} {Phys.Rev.}\ }\textbf {\bibinfo {volume}
  {D23}},\ \bibinfo {pages} {165} (\bibinfo {year} {1981})}\BibitemShut
  {NoStop}%
\bibitem [{\citenamefont {Ma}\ and\ \citenamefont {Sarkar}(1998)}]{Ma:1998dx}%
  \BibitemOpen
  \bibfield  {author} {\bibinfo {author} {\bibfnamefont {E.}~\bibnamefont
  {Ma}}\ and\ \bibinfo {author} {\bibfnamefont {U.}~\bibnamefont {Sarkar}},\
  }\href {\doibase 10.1103/PhysRevLett.80.5716} {\bibfield  {journal} {\bibinfo
   {journal} {Phys.Rev.Lett.}\ }\textbf {\bibinfo {volume} {80}},\ \bibinfo
  {pages} {5716} (\bibinfo {year} {1998})},\ \Eprint
  {http://arxiv.org/abs/hep-ph/9802445} {arXiv:hep-ph/9802445 [hep-ph]}
  \BibitemShut {NoStop}%
\bibitem [{\citenamefont {Drees}\ \emph {et~al.}()\citenamefont {Drees},
  \citenamefont {Godbole},\ and\ \citenamefont {Roy}}]{Drees2}%
  \BibitemOpen
  \bibfield  {author} {\bibinfo {author} {\bibfnamefont {M.}~\bibnamefont
  {Drees}}, \bibinfo {author} {\bibfnamefont {R.}~\bibnamefont {Godbole}}, \
  and\ \bibinfo {author} {\bibfnamefont {P.}~\bibnamefont {Roy}},\ }\href@noop
  {} {\emph {\bibinfo {title} {{Theory and phenomenology of sparticles: An
  account of four-dimensional N=1 supersymmetry in high energy physics}}}},\
  \bibinfo {note} {hackensack, USA: World Scientific (2004) 555 p}\BibitemShut
  {NoStop}%
\bibitem [{\citenamefont {Falk}\ \emph {et~al.}(1994)\citenamefont {Falk},
  \citenamefont {Olive},\ and\ \citenamefont {Srednicki}}]{Falk:1994es}%
  \BibitemOpen
  \bibfield  {author} {\bibinfo {author} {\bibfnamefont {T.}~\bibnamefont
  {Falk}}, \bibinfo {author} {\bibfnamefont {K.~A.}\ \bibnamefont {Olive}}, \
  and\ \bibinfo {author} {\bibfnamefont {M.}~\bibnamefont {Srednicki}},\ }\href
  {\doibase 10.1016/0370-2693(94)90639-4} {\bibfield  {journal} {\bibinfo
  {journal} {Phys.Lett.}\ }\textbf {\bibinfo {volume} {B339}},\ \bibinfo
  {pages} {248} (\bibinfo {year} {1994})},\ \Eprint
  {http://arxiv.org/abs/hep-ph/9409270} {arXiv:hep-ph/9409270 [hep-ph]}
  \BibitemShut {NoStop}%
\bibitem [{\citenamefont {Borzumati}\ \emph {et~al.}(1996)\citenamefont
  {Borzumati}, \citenamefont {Grossman}, \citenamefont {Nardi},\ and\
  \citenamefont {Nir}}]{Borzumati:1996hd}%
  \BibitemOpen
  \bibfield  {author} {\bibinfo {author} {\bibfnamefont {F.}~\bibnamefont
  {Borzumati}}, \bibinfo {author} {\bibfnamefont {Y.}~\bibnamefont {Grossman}},
  \bibinfo {author} {\bibfnamefont {E.}~\bibnamefont {Nardi}}, \ and\ \bibinfo
  {author} {\bibfnamefont {Y.}~\bibnamefont {Nir}},\ }\href {\doibase
  10.1016/0370-2693(96)00823-4} {\bibfield  {journal} {\bibinfo  {journal}
  {Phys.Lett.}\ }\textbf {\bibinfo {volume} {B384}},\ \bibinfo {pages} {123}
  (\bibinfo {year} {1996})},\ \Eprint {http://arxiv.org/abs/hep-ph/9606251}
  {arXiv:hep-ph/9606251 [hep-ph]} \BibitemShut {NoStop}%
\bibitem [{\citenamefont {Diaz}\ \emph {et~al.}(1998)\citenamefont {Diaz},
  \citenamefont {Romao},\ and\ \citenamefont {Valle}}]{Diaz:1997xc}%
  \BibitemOpen
  \bibfield  {author} {\bibinfo {author} {\bibfnamefont {M.~A.}\ \bibnamefont
  {Diaz}}, \bibinfo {author} {\bibfnamefont {J.~C.}\ \bibnamefont {Romao}}, \
  and\ \bibinfo {author} {\bibfnamefont {J.}~\bibnamefont {Valle}},\ }\href
  {\doibase 10.1016/S0550-3213(98)00234-X} {\bibfield  {journal} {\bibinfo
  {journal} {Nucl.Phys.}\ }\textbf {\bibinfo {volume} {B524}},\ \bibinfo
  {pages} {23} (\bibinfo {year} {1998})},\ \Eprint
  {http://arxiv.org/abs/hep-ph/9706315} {arXiv:hep-ph/9706315 [hep-ph]}
  \BibitemShut {NoStop}%
\bibitem [{\citenamefont {Barbier}\ \emph {et~al.}(2005)\citenamefont
  {Barbier}, \citenamefont {Berat}, \citenamefont {Besancon}, \citenamefont
  {Chemtob}, \citenamefont {Deandrea} \emph {et~al.}}]{Barbier:2004ez}%
  \BibitemOpen
  \bibfield  {author} {\bibinfo {author} {\bibfnamefont {R.}~\bibnamefont
  {Barbier}}, \bibinfo {author} {\bibfnamefont {C.}~\bibnamefont {Berat}},
  \bibinfo {author} {\bibfnamefont {M.}~\bibnamefont {Besancon}}, \bibinfo
  {author} {\bibfnamefont {M.}~\bibnamefont {Chemtob}}, \bibinfo {author}
  {\bibfnamefont {A.}~\bibnamefont {Deandrea}},  \emph {et~al.},\ }\href
  {\doibase 10.1016/j.physrep.2005.08.006} {\bibfield  {journal} {\bibinfo
  {journal} {Phys.Rept.}\ }\textbf {\bibinfo {volume} {420}},\ \bibinfo {pages}
  {1} (\bibinfo {year} {2005})},\ \Eprint {http://arxiv.org/abs/hep-ph/0406039}
  {arXiv:hep-ph/0406039 [hep-ph]} \BibitemShut {NoStop}%
\bibitem [{\citenamefont {Bhattacharyya}(1997)}]{Bhattacharyya:1996nj}%
  \BibitemOpen
  \bibfield  {author} {\bibinfo {author} {\bibfnamefont {G.}~\bibnamefont
  {Bhattacharyya}},\ }\href {\doibase 10.1016/S0920-5632(96)00539-7} {\bibfield
   {journal} {\bibinfo  {journal} {Nucl.Phys.Proc.Suppl.}\ }\textbf {\bibinfo
  {volume} {52A}},\ \bibinfo {pages} {83} (\bibinfo {year} {1997})},\ \Eprint
  {http://arxiv.org/abs/hep-ph/9608415} {arXiv:hep-ph/9608415 [hep-ph]}
  \BibitemShut {NoStop}%
\bibitem [{\citenamefont {Ma}\ and\ \citenamefont {Sarkar}(2012)}]{Ma:2011zm}%
  \BibitemOpen
  \bibfield  {author} {\bibinfo {author} {\bibfnamefont {E.}~\bibnamefont
  {Ma}}\ and\ \bibinfo {author} {\bibfnamefont {U.}~\bibnamefont {Sarkar}},\
  }\href {\doibase 10.1103/PhysRevD.85.075015} {\bibfield  {journal} {\bibinfo
  {journal} {Phys.Rev.}\ }\textbf {\bibinfo {volume} {D85}},\ \bibinfo {pages}
  {075015} (\bibinfo {year} {2012})},\ \Eprint {http://arxiv.org/abs/1111.5350}
  {arXiv:1111.5350 [hep-ph]} \BibitemShut {NoStop}%
\bibitem [{\citenamefont {Hall}\ \emph {et~al.}(1998)\citenamefont {Hall},
  \citenamefont {Moroi},\ and\ \citenamefont {Murayama}}]{Hall:1997ah}%
  \BibitemOpen
  \bibfield  {author} {\bibinfo {author} {\bibfnamefont {L.~J.}\ \bibnamefont
  {Hall}}, \bibinfo {author} {\bibfnamefont {T.}~\bibnamefont {Moroi}}, \ and\
  \bibinfo {author} {\bibfnamefont {H.}~\bibnamefont {Murayama}},\ }\href
  {\doibase 10.1016/S0370-2693(98)00196-8} {\bibfield  {journal} {\bibinfo
  {journal} {Phys.Lett.}\ }\textbf {\bibinfo {volume} {B424}},\ \bibinfo
  {pages} {305} (\bibinfo {year} {1998})},\ \Eprint
  {http://arxiv.org/abs/hep-ph/9712515} {arXiv:hep-ph/9712515 [hep-ph]}
  \BibitemShut {NoStop}%
\bibitem [{\citenamefont {Tucker-Smith}\ and\ \citenamefont
  {Weiner}(2001)}]{TuckerSmith:2001hy}%
  \BibitemOpen
  \bibfield  {author} {\bibinfo {author} {\bibfnamefont {D.}~\bibnamefont
  {Tucker-Smith}}\ and\ \bibinfo {author} {\bibfnamefont {N.}~\bibnamefont
  {Weiner}},\ }\href {\doibase 10.1103/PhysRevD.64.043502} {\bibfield
  {journal} {\bibinfo  {journal} {Phys.Rev.}\ }\textbf {\bibinfo {volume}
  {D64}},\ \bibinfo {pages} {043502} (\bibinfo {year} {2001})},\ \Eprint
  {http://arxiv.org/abs/hep-ph/0101138} {arXiv:hep-ph/0101138 [hep-ph]}
  \BibitemShut {NoStop}%
\bibitem [{\citenamefont {Tucker-Smith}\ and\ \citenamefont
  {Weiner}(2005)}]{TuckerSmith:2004jv}%
  \BibitemOpen
  \bibfield  {author} {\bibinfo {author} {\bibfnamefont {D.}~\bibnamefont
  {Tucker-Smith}}\ and\ \bibinfo {author} {\bibfnamefont {N.}~\bibnamefont
  {Weiner}},\ }\href {\doibase 10.1103/PhysRevD.72.063509} {\bibfield
  {journal} {\bibinfo  {journal} {Phys.Rev.}\ }\textbf {\bibinfo {volume}
  {D72}},\ \bibinfo {pages} {063509} (\bibinfo {year} {2005})},\ \Eprint
  {http://arxiv.org/abs/hep-ph/0402065} {arXiv:hep-ph/0402065 [hep-ph]}
  \BibitemShut {NoStop}%
\bibitem [{\citenamefont {Bernabei}\ \emph {et~al.}(2013)\citenamefont
  {Bernabei}, \citenamefont {Belli}, \citenamefont {Cappella}, \citenamefont
  {Caracciolo}, \citenamefont {Castellano} \emph {et~al.}}]{Bernabei:2013xsa}%
  \BibitemOpen
  \bibfield  {author} {\bibinfo {author} {\bibfnamefont {R.}~\bibnamefont
  {Bernabei}}, \bibinfo {author} {\bibfnamefont {P.}~\bibnamefont {Belli}},
  \bibinfo {author} {\bibfnamefont {F.}~\bibnamefont {Cappella}}, \bibinfo
  {author} {\bibfnamefont {V.}~\bibnamefont {Caracciolo}}, \bibinfo {author}
  {\bibfnamefont {S.}~\bibnamefont {Castellano}},  \emph {et~al.},\ }\href
  {\doibase 10.1140/epjc/s10052-013-2648-7} {\bibfield  {journal} {\bibinfo
  {journal} {Eur.Phys.J.}\ }\textbf {\bibinfo {volume} {C73}},\ \bibinfo
  {pages} {2648} (\bibinfo {year} {2013})},\ \Eprint
  {http://arxiv.org/abs/1308.5109} {arXiv:1308.5109 [astro-ph.GA]} \BibitemShut
  {NoStop}%
\bibitem [{\citenamefont {Aprile}\ \emph {et~al.}(2011)\citenamefont {Aprile}
  \emph {et~al.}}]{Aprile:2011ts}%
  \BibitemOpen
  \bibfield  {author} {\bibinfo {author} {\bibfnamefont {E.}~\bibnamefont
  {Aprile}} \emph {et~al.} (\bibinfo {collaboration} {XENON100
  Collaboration}),\ }\href {\doibase 10.1103/PhysRevD.84.061101} {\bibfield
  {journal} {\bibinfo  {journal} {Phys.Rev.}\ }\textbf {\bibinfo {volume}
  {D84}},\ \bibinfo {pages} {061101} (\bibinfo {year} {2011})},\ \Eprint
  {http://arxiv.org/abs/1104.3121} {arXiv:1104.3121 [astro-ph.CO]} \BibitemShut
  {NoStop}%
\bibitem [{\citenamefont {Arina}\ \emph {et~al.}(2013)\citenamefont {Arina},
  \citenamefont {Mohapatra},\ and\ \citenamefont {Sahu}}]{Arina:2012aj}%
  \BibitemOpen
  \bibfield  {author} {\bibinfo {author} {\bibfnamefont {C.}~\bibnamefont
  {Arina}}, \bibinfo {author} {\bibfnamefont {R.~N.}\ \bibnamefont
  {Mohapatra}}, \ and\ \bibinfo {author} {\bibfnamefont {N.}~\bibnamefont
  {Sahu}},\ }\href {\doibase 10.1016/j.physletb.2013.01.059} {\bibfield
  {journal} {\bibinfo  {journal} {Phys.Lett.}\ }\textbf {\bibinfo {volume}
  {B720}},\ \bibinfo {pages} {130} (\bibinfo {year} {2013})},\ \Eprint
  {http://arxiv.org/abs/1211.0435} {arXiv:1211.0435 [hep-ph]} \BibitemShut
  {NoStop}%
\bibitem [{\citenamefont {Grossman}\ and\ \citenamefont
  {Haber}(1997)}]{Grossman:1997is}%
  \BibitemOpen
  \bibfield  {author} {\bibinfo {author} {\bibfnamefont {Y.}~\bibnamefont
  {Grossman}}\ and\ \bibinfo {author} {\bibfnamefont {H.~E.}\ \bibnamefont
  {Haber}},\ }\href {\doibase 10.1103/PhysRevLett.78.3438} {\bibfield
  {journal} {\bibinfo  {journal} {Phys.Rev.Lett.}\ }\textbf {\bibinfo {volume}
  {78}},\ \bibinfo {pages} {3438} (\bibinfo {year} {1997})},\ \Eprint
  {http://arxiv.org/abs/hep-ph/9702421} {arXiv:hep-ph/9702421 [hep-ph]}
  \BibitemShut {NoStop}%
\bibitem [{\citenamefont {Arina}\ and\ \citenamefont
  {Sahu}(2012)}]{Arina:2011cu}%
  \BibitemOpen
  \bibfield  {author} {\bibinfo {author} {\bibfnamefont {C.}~\bibnamefont
  {Arina}}\ and\ \bibinfo {author} {\bibfnamefont {N.}~\bibnamefont {Sahu}},\
  }\href {\doibase 10.1016/j.nuclphysb.2011.09.014} {\bibfield  {journal}
  {\bibinfo  {journal} {Nucl.Phys.}\ }\textbf {\bibinfo {volume} {B854}},\
  \bibinfo {pages} {666} (\bibinfo {year} {2012})},\ \Eprint
  {http://arxiv.org/abs/1108.3967} {arXiv:1108.3967 [hep-ph]} \BibitemShut
  {NoStop}%
\bibitem [{\citenamefont {Buckley}\ and\ \citenamefont
  {Profumo}(2012)}]{Buckley:2011ye}%
  \BibitemOpen
  \bibfield  {author} {\bibinfo {author} {\bibfnamefont {M.~R.}\ \bibnamefont
  {Buckley}}\ and\ \bibinfo {author} {\bibfnamefont {S.}~\bibnamefont
  {Profumo}},\ }\href {\doibase 10.1103/PhysRevLett.108.011301} {\bibfield
  {journal} {\bibinfo  {journal} {Phys.Rev.Lett.}\ }\textbf {\bibinfo {volume}
  {108}},\ \bibinfo {pages} {011301} (\bibinfo {year} {2012})},\ \Eprint
  {http://arxiv.org/abs/1109.2164} {arXiv:1109.2164 [hep-ph]} \BibitemShut
  {NoStop}%
\bibitem [{\citenamefont {Cirelli}\ \emph {et~al.}(2012)\citenamefont
  {Cirelli}, \citenamefont {Panci}, \citenamefont {Servant},\ and\
  \citenamefont {Zaharijas}}]{Cirelli:2011ac}%
  \BibitemOpen
  \bibfield  {author} {\bibinfo {author} {\bibfnamefont {M.}~\bibnamefont
  {Cirelli}}, \bibinfo {author} {\bibfnamefont {P.}~\bibnamefont {Panci}},
  \bibinfo {author} {\bibfnamefont {G.}~\bibnamefont {Servant}}, \ and\
  \bibinfo {author} {\bibfnamefont {G.}~\bibnamefont {Zaharijas}},\ }\href
  {\doibase 10.1088/1475-7516/2012/03/015} {\bibfield  {journal} {\bibinfo
  {journal} {JCAP}\ }\textbf {\bibinfo {volume} {1203}},\ \bibinfo {pages}
  {015} (\bibinfo {year} {2012})},\ \Eprint {http://arxiv.org/abs/1110.3809}
  {arXiv:1110.3809 [hep-ph]} \BibitemShut {NoStop}%
\bibitem [{\citenamefont {Hambye}\ \emph {et~al.}(2001)\citenamefont {Hambye},
  \citenamefont {Ma},\ and\ \citenamefont {Sarkar}}]{Hambye:2000ui}%
  \BibitemOpen
  \bibfield  {author} {\bibinfo {author} {\bibfnamefont {T.}~\bibnamefont
  {Hambye}}, \bibinfo {author} {\bibfnamefont {E.}~\bibnamefont {Ma}}, \ and\
  \bibinfo {author} {\bibfnamefont {U.}~\bibnamefont {Sarkar}},\ }\href
  {\doibase 10.1016/S0550-3213(01)00109-2} {\bibfield  {journal} {\bibinfo
  {journal} {Nucl.Phys.}\ }\textbf {\bibinfo {volume} {B602}},\ \bibinfo
  {pages} {23} (\bibinfo {year} {2001})},\ \Eprint
  {http://arxiv.org/abs/hep-ph/0011192} {arXiv:hep-ph/0011192 [hep-ph]}
  \BibitemShut {NoStop}%
\bibitem [{\citenamefont {Chun}\ and\ \citenamefont
  {Scopel}(2007)}]{Chun:2006sp}%
  \BibitemOpen
  \bibfield  {author} {\bibinfo {author} {\bibfnamefont {E.~J.}\ \bibnamefont
  {Chun}}\ and\ \bibinfo {author} {\bibfnamefont {S.}~\bibnamefont {Scopel}},\
  }\href {\doibase 10.1103/PhysRevD.75.023508} {\bibfield  {journal} {\bibinfo
  {journal} {Phys.Rev.}\ }\textbf {\bibinfo {volume} {D75}},\ \bibinfo {pages}
  {023508} (\bibinfo {year} {2007})},\ \Eprint
  {http://arxiv.org/abs/hep-ph/0609259} {arXiv:hep-ph/0609259 [hep-ph]}
  \BibitemShut {NoStop}%
\bibitem [{\citenamefont {Hagelin}\ \emph {et~al.}(1984)\citenamefont
  {Hagelin}, \citenamefont {Kane},\ and\ \citenamefont
  {Raby}}]{Hagelin:1984wv}%
  \BibitemOpen
  \bibfield  {author} {\bibinfo {author} {\bibfnamefont {J.~S.}\ \bibnamefont
  {Hagelin}}, \bibinfo {author} {\bibfnamefont {G.~L.}\ \bibnamefont {Kane}}, \
  and\ \bibinfo {author} {\bibfnamefont {S.}~\bibnamefont {Raby}},\ }\href
  {\doibase 10.1016/0550-3213(84)90064-6} {\bibfield  {journal} {\bibinfo
  {journal} {Nucl.Phys.}\ }\textbf {\bibinfo {volume} {B241}},\ \bibinfo
  {pages} {638} (\bibinfo {year} {1984})}\BibitemShut {NoStop}%
\bibitem [{\citenamefont {Belanger}\ \emph {et~al.}(2006)\citenamefont
  {Belanger}, \citenamefont {Boudjema}, \citenamefont {Pukhov},\ and\
  \citenamefont {Semenov}}]{Belanger:2004yn}%
  \BibitemOpen
  \bibfield  {author} {\bibinfo {author} {\bibfnamefont {G.}~\bibnamefont
  {Belanger}}, \bibinfo {author} {\bibfnamefont {F.}~\bibnamefont {Boudjema}},
  \bibinfo {author} {\bibfnamefont {A.}~\bibnamefont {Pukhov}}, \ and\ \bibinfo
  {author} {\bibfnamefont {A.}~\bibnamefont {Semenov}},\ }\href {\doibase
  10.1016/j.cpc.2005.12.005} {\bibfield  {journal} {\bibinfo  {journal}
  {Comput.Phys.Commun.}\ }\textbf {\bibinfo {volume} {174}},\ \bibinfo {pages}
  {577} (\bibinfo {year} {2006})},\ \Eprint
  {http://arxiv.org/abs/hep-ph/0405253} {arXiv:hep-ph/0405253 [hep-ph]}
  \BibitemShut {NoStop}%
\bibitem [{\citenamefont {Belanger}\ \emph {et~al.}(2014)\citenamefont
  {Belanger}, \citenamefont {Boudjema}, \citenamefont {Pukhov},\ and\
  \citenamefont {Semenov}}]{Belanger:2013oya}%
  \BibitemOpen
  \bibfield  {author} {\bibinfo {author} {\bibfnamefont {G.}~\bibnamefont
  {Belanger}}, \bibinfo {author} {\bibfnamefont {F.}~\bibnamefont {Boudjema}},
  \bibinfo {author} {\bibfnamefont {A.}~\bibnamefont {Pukhov}}, \ and\ \bibinfo
  {author} {\bibfnamefont {A.}~\bibnamefont {Semenov}},\ }\href {\doibase
  10.1016/j.cpc.2013.10.016} {\bibfield  {journal} {\bibinfo  {journal}
  {Comput.Phys.Commun.}\ }\textbf {\bibinfo {volume} {185}},\ \bibinfo {pages}
  {960} (\bibinfo {year} {2014})},\ \Eprint {http://arxiv.org/abs/1305.0237}
  {arXiv:1305.0237 [hep-ph]} \BibitemShut {NoStop}%
\bibitem [{\citenamefont {Djouadi}\ \emph {et~al.}(2007)\citenamefont
  {Djouadi}, \citenamefont {Kneur},\ and\ \citenamefont
  {Moultaka}}]{Djouadi:2002ze}%
  \BibitemOpen
  \bibfield  {author} {\bibinfo {author} {\bibfnamefont {A.}~\bibnamefont
  {Djouadi}}, \bibinfo {author} {\bibfnamefont {J.-L.}\ \bibnamefont {Kneur}},
  \ and\ \bibinfo {author} {\bibfnamefont {G.}~\bibnamefont {Moultaka}},\
  }\href {\doibase 10.1016/j.cpc.2006.11.009} {\bibfield  {journal} {\bibinfo
  {journal} {Comput.Phys.Commun.}\ }\textbf {\bibinfo {volume} {176}},\
  \bibinfo {pages} {426} (\bibinfo {year} {2007})},\ \Eprint
  {http://arxiv.org/abs/hep-ph/0211331} {arXiv:hep-ph/0211331 [hep-ph]}
  \BibitemShut {NoStop}%
\bibitem [{\citenamefont {Kolb}\ and\ \citenamefont {Turner}(1994)}]{kotu}%
  \BibitemOpen
  \bibfield  {author} {\bibinfo {author} {\bibfnamefont {E.}~\bibnamefont
  {Kolb}}\ and\ \bibinfo {author} {\bibfnamefont {M.}~\bibnamefont {Turner}},\
  }\href@noop {} {\emph {\bibinfo {title} {The Early Universe}}}\ (\bibinfo
  {publisher} {Westview Press},\ \bibinfo {year} {1994})\BibitemShut {NoStop}%
\bibitem [{\citenamefont {{Griest, Kim and Seckel, David}}(1991)}]{coann}%
  \BibitemOpen
  \bibfield  {author} {\bibinfo {author} {\bibnamefont {{Griest, Kim and
  Seckel, David}}},\ }\href {\doibase 10.1103/PhysRevD.43.3191} {\bibfield
  {journal} {\bibinfo  {journal} {Phys. Rev. D}\ }\textbf {\bibinfo {volume}
  {43}},\ \bibinfo {pages} {3191} (\bibinfo {year} {1991})}\BibitemShut
  {NoStop}%
\bibitem [{\citenamefont {Aprile}\ \emph {et~al.}(2012)\citenamefont {Aprile}
  \emph {et~al.}}]{Aprile:2012nq}%
  \BibitemOpen
  \bibfield  {author} {\bibinfo {author} {\bibfnamefont {E.}~\bibnamefont
  {Aprile}} \emph {et~al.} (\bibinfo {collaboration} {XENON100
  Collaboration}),\ }\href {\doibase 10.1103/PhysRevLett.109.181301} {\bibfield
   {journal} {\bibinfo  {journal} {Phys.Rev.Lett.}\ }\textbf {\bibinfo {volume}
  {109}},\ \bibinfo {pages} {181301} (\bibinfo {year} {2012})},\ \Eprint
  {http://arxiv.org/abs/1207.5988} {arXiv:1207.5988 [astro-ph.CO]} \BibitemShut
  {NoStop}%
\bibitem [{\citenamefont {Akerib}\ \emph {et~al.}(2013)\citenamefont {Akerib}
  \emph {et~al.}}]{Akerib:2013tjd}%
  \BibitemOpen
  \bibfield  {author} {\bibinfo {author} {\bibfnamefont {D.}~\bibnamefont
  {Akerib}} \emph {et~al.} (\bibinfo {collaboration} {LUX Collaboration}),\
  }\href@noop {} {\  (\bibinfo {year} {2013})},\ \Eprint
  {http://arxiv.org/abs/1310.8214} {arXiv:1310.8214 [astro-ph.CO]} \BibitemShut
  {NoStop}%
\bibitem [{\citenamefont {Edsjo}\ and\ \citenamefont
  {Gondolo}(1997)}]{Edsjo:1997bg}%
  \BibitemOpen
  \bibfield  {author} {\bibinfo {author} {\bibfnamefont {J.}~\bibnamefont
  {Edsjo}}\ and\ \bibinfo {author} {\bibfnamefont {P.}~\bibnamefont
  {Gondolo}},\ }\href {\doibase 10.1103/PhysRevD.56.1879} {\bibfield  {journal}
  {\bibinfo  {journal} {Phys.Rev.}\ }\textbf {\bibinfo {volume} {D56}},\
  \bibinfo {pages} {1879} (\bibinfo {year} {1997})},\ \Eprint
  {http://arxiv.org/abs/hep-ph/9704361} {arXiv:hep-ph/9704361 [hep-ph]}
  \BibitemShut {NoStop}%
\end{thebibliography}%

\end{document}